\documentclass[aps,floats]{revtex4}
\usepackage{amsmath,amssymb}
\usepackage{graphicx,epsfig}
\usepackage[greek,english]{babel}
\usepackage{bbold}

\begin{document}
\bibliographystyle {plain}

\pdfoutput=1
\def\oppropto{\mathop{\propto}} 
\def\opsimeq{\mathop{\simeq}}
\def\opoverderline{\mathop{\overline}}
\def\operarrow{\mathop{\longrightarrow}}
\def\opsim{\mathop{\sim}}

\title{ Revisiting boundary-driven non-equilibrium Markov dynamics in arbitrary potentials \\
via supersymmetric quantum mechanics and via explicit large deviations at various levels} 


\author{ C\'ecile Monthus }
 \affiliation{Universit\'e Paris Saclay, CNRS, CEA,
 Institut de Physique Th\'eorique, 
91191 Gif-sur-Yvette, France}

\begin{abstract}

For boundary-driven non-equilibrium Markov models of non-interacting particles in one dimension, either in continuous space with the Fokker-Planck dynamics involving an arbitrary force $F(x)$ and an arbitrary diffusion coefficient $D(x)$, or in discrete space with the Markov jump dynamics involving arbitrary nearest-neighbor transition rates $w(x \pm 1,x)$, the Markov generator can be transformed via an appropriate similarity transformation into a quantum supersymmetric Hamiltonian with many remarkable properties. We first describe how the mapping from the boundary-driven non-equilibrium dynamics towards some dual equilibrium dynamics [J. Tailleur, J. Kurchan and V. Lecomte, J. Phys. A 41, 505001 (2008)] can be reinterpreted via the two corresponding quantum Hamiltonians that are supersymmetric partners of each other, with the same energy spectra. We describe the consequences for the spectral decomposition of the boundary-driven dynamics, and we give explicit expressions for the Kemeny times needed to converge towards the non-equilibrium steady states. We then focus on the large deviations at various levels for empirical time-averaged observables over a large time-window $T$. We start with the always explicit Level 2.5 concerning the joint distribution of the empirical density and of the empirical flows before considering the contractions towards lower levels. In particular, the rate function for the empirical current alone can be explicitly computed via the contraction from the Level 2.5 using the properties of the associated quantum supersymmetric Hamiltonians.

\end{abstract}

\maketitle

\section{ Introduction   }

The theory of large deviations 
 (see the reviews \cite{oono,ellis,review_touchette} and references therein)
 has become the cornerstone of nonequilibrium statistical physics
(see the reviews with different scopes \cite{derrida-lecture,harris_Schu,searles,harris,mft,sollich_review,lazarescu_companion,lazarescu_generic,jack_review}, 
the PhD Theses \cite{fortelle_thesis,vivien_thesis,chetrite_thesis,wynants_thesis,chabane_thesis,duBuisson_thesis} 
 and the Habilitation Thesis \cite{chetrite_HDR}).
In particular, the large deviations of time-averaged observables over a large time-window $T$
have attracted a lot of interest for many Markov processes 
 \cite{peliti,derrida-lecture,sollich_review,lazarescu_companion,lazarescu_generic,jack_review,vivien_thesis,lecomte_chaotic,lecomte_thermo,lecomte_formalism,lecomte_glass,kristina1,kristina2,jack_ensemble,simon1,simon2,tailleur,simon3,Gunter1,Gunter2,Gunter3,Gunter4,chetrite_canonical,chetrite_conditioned,chetrite_optimal,chetrite_HDR,touchette_circle,touchette_langevin,touchette_occ,touchette_occupation,garrahan_lecture,Vivo,c_ring,c_detailed,chemical,derrida-conditioned,derrida-ring,bertin-conditioned,touchette-reflected,touchette-reflectedbis,c_lyapunov,previousquantum2.5doob,quantum2.5doob,quantum2.5dooblong,c_ruelle,lapolla,c_east,chabane,us_gyrator,duBuisson_gyrator,c_largedevpearson} with the construction of the corresponding Doob's conditioned processes.
Since the rate functions for time-averaged observables are not always explicit,
it is interesting to consider the large deviations at higher levels in order to obtain explicit rate functions
for Markov processes with arbitrary generators.
An important result is that the large deviations at Level 2 concerning
the distribution of the empirical density over a large time-window $T$
are always explicit for equilibrium Markov processes 
with vanishing steady currents satisfying detailed-balance.
However, for non-equilibrium Markov processes 
with non-vanishing steady currents breaking detailed-balance,
the Level 2 for the empirical density alone is not always explicit, and 
the Level 2.5 concerning the joint distribution of the empirical density and of the empirical flows
is the appropriate level 
where rate functions can be written explicitly in full generality, in particular for discrete-time Markov chains 
\cite{fortelle_thesis,fortelle_chain,review_touchette,c_largedevdisorder,c_reset,c_inference,c_microcanoEnsembles},
for continuous-time Markov jump processes with discrete space
\cite{fortelle_thesis,fortelle_jump,maes_canonical,maes_onandbeyond,wynants_thesis,chetrite_formal,BFG1,BFG2,chetrite_HDR,c_ring,c_interactions,c_open,barato_periodic,chetrite_periodic,c_reset,c_inference,c_LargeDevAbsorbing,c_microcanoEnsembles},
for diffusion processes in continuous space
\cite{wynants_thesis,maes_diffusion,chetrite_formal,engel,chetrite_HDR,c_lyapunov,c_inference}, 
or for jump-drift processes \cite{c_reset,c_runandtumble,c_jumpdrift,c_SkewDB}.

Among non-equilibrium dynamics, the simplest examples are boundary-driven models, 
where equilibrium steady states would be possible if the boundary conditions were different,
i.e. the non-equilibrium aspect comes only from the boundary conditions and not from the bulk dynamical rules. 
Since whenever there is detailed-balance, 
the Markov generator can be transformed via an appropriate similarity transformation
into a symmetric operator (see the textbooks \cite{gardiner,vankampen,risken}) 
that can be further interpreted as a quantum supersymmetric Hamiltonian
with many remarkable properties
(see the review \cite{review_susyquantum} and references therein),
it is interesting to make the same similarity transformation
 for boundary-driven one-dimensional Markov models of non-interacting particles
 in arbitrary potentials.
 The first goal of the present paper is to describe how the mapping from boundary-driven dynamics towards 
 some dual equilibrium dynamics (see the sections concerning non-interacting particles in \cite{tailleurMapping}) 
 can be then reinterpreted via
 quantum Hamiltonians that are supersymmetric partners, with the same energy spectra,
 while their eigenvectors are related to each other via simple formula.
 We will then analyze the consequences for the spectral decomposition
 of the boundary driven dynamics,
 as well as for the Kemeny times needed to converge towards the non-equilibrium steady states 
 (see \cite{us_kemeny} and references therein)
 in order to obtain explicit expressions for arbitrary potentials
 even if the individual relaxation eigenvalues are not explicitly known.
 
The second goal of the present paper is to study the large deviations at various levels for the boundary-driven 
one-dimensional Markov models of non-interacting particles in arbitrary potentials.
 We will start from the always explicit rate functions at Level 2.5.
 Indeed, while the Level 2.5 is usually written for a single Markov trajectory or
 for closed systems with a fixed number of particles,
 the generalization towards open systems in contact with particle reservoirs is straightforward \cite{c_open}.
 We will then analyze the contractions towards lower levels,
 i.e. the optimization over the empirical observables that one wishes to integrate out from the Level 2.5,
 in order to see if the rate functions of the remaining empirical observables
 can be explicitly written.
 In particular, we will compute the rate function for the empirical current alone
via the contraction from the Level 2.5, where the quantum supersymmetric Hamiltonians
play an essential role, as already found for non-equilibrium diffusion processes on a ring
with periodic boundary conditions \cite{c_lyapunov}.

For clarity, the present paper is divided into two separate parts :
the main text is entirely devoted to the boundary-driven non-equilibrium 
Fokker-Planck dynamics with an arbitrary force $F(x)$ and an arbitrary diffusion coefficient $D(x)$
on the continuous spatial interval $x \in [0,L]$, 
while the various Appendices describe the analog results for the Markov jump dynamics 
on the discrete sites $x =0,1,..,L$ with arbitrary nearest-neighbor transition rates $w(x \pm 1,x)$:
the physics is the same, but there are some important technical differences 
that are interesting to stress.

The main text is organized as follows.
In section \ref{sec_boundaryFP}, 
we introduce the boundary-driven non-equilibrium 
Fokker-Planck dynamics and recall the mapping towards 
 some dual equilibrium dynamics \cite{tailleurMapping}.
In section \ref{sec_susyFP}, this mapping is reinterpreted
via the corresponding quantum Hamiltonians
 that are supersymmetric partners of each other.
In section \ref{sec_spectralFP}, we use the spectral properties of the quantum 
supersymmetric Hamiltonians to analyze the spectral decompositions of the
Fokker-Planck dynamics.
In section \ref{sec_kemenyFP}, we give the explicit expression
for the Kemeny time needed to converge towards the non-equilibrium steady state
and we describe an illustrative example involving a saw-tooth potential
that can be either a valley or a mountain in order to discuss the scaling of the Kemeny time with the system size $L$.
In section \ref{sec_largedevFP}, 
we turn to the analysis of the large deviations properties
at various levels. We start with the always explicit Level 2.5
concerning the joint distribution of the empirical density and of the empirical current,
before considering the contractions towards lower levels : 
we write the explicit rate function at Level 2 concerning the empirical density alone,
as well as the explicit rate function for the empirical current alone.
Our conclusions are summarized in section \ref{sec_conclusion}.


\section{ Reminder on boundary-driven non-equilibrium Fokker-Planck dynamics  }

\label{sec_boundaryFP}

In this section, we describe the boundary-driven non-equilibrium Fokker-Planck dynamics with an arbitrary force $F(x)$ and an arbitrary diffusion coefficient $D(x)$
and we recall the mapping towards some dual equilibrium dynamics \cite{tailleurMapping}.

\subsection{ Fokker-Planck non-equilibrium dynamics between two reservoirs fixing the boundary densities  }

\subsubsection{ Dynamics involving the Fokker-Planck force $F(x)$, the diffusion coefficient $D(x)$ and the corresponding potential $U(x)$ }

The Fokker-Planck dynamics for the density $ \rho_t( x ) $ on the interval $0 \leq x \leq L$  
is written as the continuity equation
\begin{eqnarray}
 \partial_t \rho_t( x )    =  - \partial_x J_t(x)
\label{fokkerplanck}
\end{eqnarray}
where the current $J_t(x) $ associated with the density $\rho_t(x) $
involves the Fokker-Planck force $F(x) $ and the diffusion coefficient $D(x)$
\begin{eqnarray}
 J_t( x ) =F(x)  \rho_t(x)   - D(x) \partial_x  \rho_t(x)
   = - D(x) \left[ U'(x) + \partial_x \right] \rho_t(x)  
\label{fokkerplanckcurrent}
\end{eqnarray}
where in the last expression, the force $F(x)$ has been replaced by the potential $U(x)$ 
\begin{eqnarray}
U(x) && \equiv - \int_0^x dy \frac{F(y)}{D(y)} 
\nonumber \\
U'(x) && = - \frac{F(x)}{D(x)}
 \label{Ux}
\end{eqnarray}
So the Fokker-Planck dynamics of Eq. \eqref{fokkerplanck}
\begin{eqnarray}
 \partial_t \rho_t( x )    =    {\cal L} \rho_t(x)
\label{fokkerplanckgene}
\end{eqnarray}
involves the generator
\begin{eqnarray}
  {\cal L}   =  \partial_x \left[ - F(x)    + D(x) \partial_x\right]  =  \partial_x D(x) \left[ U'(x) + \partial_x \right] 
\label{generator}
\end{eqnarray}
The boundary conditions at $x=0$ and at $x=L$ correspond to fixed densities $c_0$ and $c_L$
imposed by the external reservoirs
\begin{eqnarray}
  \rho_t( x=0 )   && = c_0
  \nonumber \\
  \rho_t(x=L) && = c_L
\label{fokkerplanckreservoirs}
\end{eqnarray}
Via Eq. \eqref{fokkerplanck}, one obtains that 
the current spatial derivative $\partial_x J_t(x) $
vanishes at the two boundaries $x=0$ and $x=L$
\begin{eqnarray}
 \partial_x J_t(x) \vert_{x=0} && =0 
 \nonumber \\
  \partial_x J_t(x) \vert_{x=L} && =0  
\label{BCdxJ}
\end{eqnarray}
The total number $N_t$ of particles on the interval $0 \leq x \leq L$
\begin{eqnarray}
 N_t \equiv \int_0^L dx \rho_t( x )   
\label{Nt}
\end{eqnarray}
follows the dynamics
\begin{eqnarray}
 \partial_t N_t \equiv \int_0^L dx \partial_t \rho_t( x )   = - \int_0^L dx \partial_x J_t( x )
 = J_t(0) - J_t(L)
\label{Ntdyn}
\end{eqnarray}
and can thus increase and decrease according to
the values of the currents $J_t(0)$ and $J_t(L)$ at the two boundaries.


\subsubsection{ Reminder on the steady state $\rho_*(x) $ and the steady current $J_*$ as a function of the boundary densities $c_0$ and $c_L$}

The steady state $\rho_*(x)$ of the Fokker-Planck dynamics of Eq. \eqref{fokkerplanck} 
can be found by requiring that the corresponding
steady current $J_*(x)=J_*$ does depend on $x$ 
\begin{eqnarray}
 J_*   = - D(x) \left[ U'(x) + \partial_x \right] \rho_*(x)
\label{fokkerplanckcurrentJsteady}
\end{eqnarray}
This first-order equation for $\rho_*(x) $ can be solved 
 using the boundary condition of Eq. \eqref{fokkerplanckreservoirs} at $x=0$
\begin{eqnarray}
  \rho_*(x) = e^{-U(x) } \left[  c_0 - J_*\int_0^x dy \frac{ e^{U(y) }  } { D(y) } \right]
\label{variationctesol}
\end{eqnarray}
while the boundary condition of Eq. \eqref{fokkerplanckreservoirs} at $x=L$
\begin{eqnarray}
   c_L = \rho_*( x=L )    =e^{-U(L) } \left[  c_0 - J_*\int_0^L dy \frac{ e^{U(y) }  } { D(y) } \right]
\label{fokkerplanckreservoirsimpose}
\end{eqnarray}
determines the steady current
\begin{eqnarray}
   J_* =  \frac{ c_0  - c_Le^{U(L) }  }{ \int_0^L dy \frac{ e^{U(y) }  } { D(y) } }
\label{solusteadycurrent}
\end{eqnarray}

In summary, the discussion is as follows :

(i) if the boundary densities $(c_0,c_L)$ imposed by the reservoirs 
satisfy $c_0  = c_Le^{U(L) }$,
then the steady current of Eq. \eqref{solusteadycurrent} vanishes $J_*= 0$,
i.e. detailed-balance is satisfied and the corresponding steady state of Eq. \eqref{variationctesol} 
reduces to the Boltzmann equilibrium in the potential $U(x)$
\begin{eqnarray}
  \rho^{eq}_*(x) = c_0 e^{  -U(x) } = c_L  e^{ U(L) -U(x) }\ \ \ {\rm if } \ \  J_* =0
\label{steadyeq}
\end{eqnarray}

(ii)  if the boundary densities $(c_0,c_L)$ imposed by the reservoirs 
correspond to a non-vanishing steady current $J_*\ne 0$ in Eq. \eqref{solusteadycurrent},
then the steady state of Eq. \eqref{variationctesol}
can be rewritten in the obviously positive expression
\begin{eqnarray}
  \rho^{noneq}_*(x) = e^{-U(x)}    \frac{  c_0 \int_x^L dy \frac{ e^{U(y) }  } { D(y) } + c_Le^{U(L) } \int_0^x dy \frac{ e^{U(y) }  } { D(y) } }{ \int_0^L dz \frac{ e^{U(z) }  } { D(z) } } 
\label{steadynoneq}
\end{eqnarray}


\subsection{ Mapping from the non-equilibrium dynamics for $\rho_t(x)$ towards some
dual equilibrium dynamics}

\subsubsection{ Reminder on the mapping of \cite{tailleurMapping} from the original density $\rho_t(x)$ 
towards another density ${\hat \rho}_t( x ) $ }

The mapping discussed in section 1 of \cite{tailleurMapping} for the special case of the constant diffusion coefficient $D(x)=T$
determined by the temperature can be written for an arbitrary space-dependent diffusion coefficient $D(x)$
as follows. The mapping from the original density $\rho_t( x ) $ towards the new density
\begin{eqnarray}
  {\hat \rho}_t( x )  = - \partial_x \left[ e^{ U(x) }   \rho_t( x ) \right]
\label{upsilonfp}
\end{eqnarray}
can be inverted via integration
\begin{eqnarray}
 \int_0^y dx {\hat \rho}_t( x ) =    \rho_t( 0 )   -  e^{ U(y) } \rho_t( y )  
 =    c_0   -  e^{ U(y) } \rho_t( y )
\label{iniPfromhatP}
\end{eqnarray}
so that the corresponding total number ${\hat N}_t $ of particles on the interval $0 \leq x \leq L$ is 
independent of time
\begin{eqnarray}
{\hat N}_t \equiv \int_0^L  dx  {\hat \rho}_t( x )  = c_0 - e^{U(L)} c_L \equiv {\hat N} 
\label{hatNt}
\end{eqnarray}
The original current $J_t(x)$ of Eq. \eqref{fokkerplanckcurrent}
 is actually proportional to the new density ${\hat \rho}_t( x ) $
\begin{eqnarray}
 J_t( x )    = - D(x) \left[ U'(x) + \partial_x \right] \rho_t(x)
 = - D(x) e^{- U(x) }  \partial_x  \left[ e^{ U(x) }   \rho_t( x )\right]  = D(x) e^{- U(x) }   {\hat \rho}_t( x )
\label{fokkerplanckcurrentmap}
\end{eqnarray}

The continuity equation satisfied by the new density of Eq. \eqref{upsilonfp}
using Eq. \eqref{fokkerplanck}
\begin{eqnarray}
 \partial_t   {\hat \rho}_t( x ) &&   
 = - \partial_x \left[ e^{ U(x) }   \partial_t \rho_t( x )\right] 
  =   \partial_x \left( e^{ U(x) }   \partial_x  J_t(x)  \right] 
 \equiv  - \partial_x  {\hat J}_{t}( x )
\label{fokkerplanckupsilon}
\end{eqnarray}
involves the new current $ {\hat J}_{t}( x ) $ 
\begin{eqnarray}
 {\hat J}_t( x ) = - e^{U(x)}  \partial_x  J_t( x ) 
\label{fokkerplanckcalJmap}
\end{eqnarray}
which is proportional to the divergence $\partial_x  J_t( x )  $
of the original current $J_t( x )  $, 
so that it vanishes at the two boundaries as a consequence of Eq. \eqref{BCdxJ}
\begin{eqnarray}
 {\hat J}_t( x=0 )&& =0 
 \nonumber \\
 {\hat J}_t( x=L ) && =0  
\label{BCdxJhat}
\end{eqnarray}
in consistency with the conservation of the number of particles ${\hat N}_t ={\hat N}$ in Eq. \eqref{hatNt}.

Plugging the current $J_t( x ) $ of Eq. \eqref{fokkerplanckcurrentmap} into Eq. \eqref{fokkerplanckcalJmap}
yields the new current ${\hat J}_{t}( x ) $ in terms of the new density ${\hat \rho}_t( x ) $
\begin{eqnarray}
 {\hat J}_{t}( x ) && = - e^{U(x)}  \partial_x \left[  e^{- U(x) }  D(x) {\hat \rho}_t( x ) \right]  
 = -    \left[ -U'(x)+\partial_x  \right]  D(x) {\hat \rho}_t( x )  
\nonumber \\
&& \equiv {\hat F}(x) {\hat \rho}_t( x )  -D(x) \partial_x {\hat \rho}_t( x ) 
\label{fokkerplanckcalJ}
\end{eqnarray}
where the diffusion coefficient $D(x)$ is the same as in the original model,
 while the original force $F(x)$ has been replaced by
 the new force  
\begin{eqnarray}
{\hat F}(x)=D(x) U'(x) - D'(x) = - F(x) - D'(x)
\label{forcehat}
\end{eqnarray}
with the corresponding new potential $  {\hat U}(x) $ analogous to Eq. \eqref{Ux} 
 \begin{eqnarray}
 {\hat U}'(x) && = - \frac{{\hat F}(x)}{D(x)} = \frac{F(x) + D'(x) }{D(x)} = - U'(x) + \frac{ D'(x) }{D(x)}
 \nonumber \\
{\hat U}(x) && =  - U(x) + \ln D(x)
 \label{Uxhat}
\end{eqnarray}

In summary, the new density ${\hat \rho}_t( x ) $ satisfies the Fokker-Planck dynamics 
\begin{eqnarray}
 \partial_t {\hat \rho}_t( x )   =  - \partial_x  {\hat J}_{t}( x ) =   {\hat {\cal L}} {\hat \rho}_t(x)
\label{fokkerplanckhat}
\end{eqnarray}
where the new generator  ${\hat {\cal L}}$ analogous to Eq. \eqref{generator}
involves the new force ${\hat F}(x) $ of Eq. \eqref{forcehat}
or equivalently the new potential ${\hat U}(x)  $ of Eq. \eqref{Uxhat}
\begin{eqnarray}
{\hat {\cal L}}   
=  \partial_x \left[ -{\hat F}(x)   + D(x) \partial_x\right]  =  \partial_x D(x) \left[  {\hat U}'(x) + \partial_x \right] 
\label{generatorhat}
\end{eqnarray}
The zero-current boundary conditions of Eq. \eqref{BCdxJhat}
translate for the density ${\hat \rho}_t( x ) $ into
\begin{eqnarray}
0 && =  {\hat J}_t( x=0 ) 
= - D(0) \left[ {\hat U}'(0)  {\hat \rho}_t( 0 )  + \partial_x {\hat \rho}_t( x ) \vert_{x=0} \right]
 \nonumber \\
0 && =  {\hat J}_t( x=L )  
=   - D(L) \left[ {\hat U}'(L)  {\hat \rho}_t( L )  + \partial_x {\hat \rho}_t( x ) \vert_{x=L}\right]
\label{BChatrho}
\end{eqnarray}


\subsubsection{ Equilibrium steady state ${\hat \rho}_*( x ) $ with vanishing steady current ${\hat J}_* =0$ for the new model }

The new steady current ${\hat J}_*( x ) $ of Eq. \eqref{fokkerplanckcalJmap} identically vanishes
\begin{eqnarray}
 {\hat J}_*( x ) = - e^{U(x)}  \partial_x  J_* =0 
\label{fokkerplanckcalJmapsteady}
\end{eqnarray}
Via the change of variables of Eq. \eqref{upsilonfp}, 
the non-equilibrium steady state $ \rho^{noneq}_*(x) $ of Eq. \eqref{steadynoneq}
is mapped onto the new steady state using $ {\hat N}= c_0 - e^{U(L)} c_L $
of Eq. \eqref{hatNt}
\begin{eqnarray}
  {\hat \rho}^{eq}_*( x )  && = - \partial_x \left[ e^{ U(x) }   \rho^{noneq}_*( x ) \right]
  =  - \partial_x \left[ \frac{  c_0 \int_x^L dy \frac{ e^{U(y) }  } { D(y) } + c_Le^{U(L) } \int_0^x dy \frac{ e^{U(y) }  } { D(y) } }{ \int_0^L dz \frac{ e^{U(z) }  } { D(z) } }  \right]
  \nonumber \\
  && = \left( \frac{  c_0 - c_Le^{U(L) }  }
  { \int_0^L dz \frac{ e^{U(z) }  } { D(z) } } \right) \frac{ e^{U(x) }  } { D(x) }
  = {\hat N} \frac{\frac{ e^{U(x) }  } { D(x) }}{ \hat Z}
  = {\hat N} \frac{ e^{- {\hat U} (x) } }{ \hat Z}
\label{hatsteady}
\end{eqnarray}
that corresponds to the Boltzmann equilibrium in the dual potential $ {\hat U}(x) $ of Eq. \eqref{Uxhat}
with the corresponding partition function
\begin{eqnarray}
\hat Z  \equiv  \int_0^L dy   e^{ -  {\hat U} (y) } = \int_0^L dy \frac{ e^{U(y) }  } { D(y) }
 \label{zhat}
\end{eqnarray}


\subsubsection{ Iterated mapping from the new density ${\hat \rho}_t( x ) $ towards another density ${\hat {\hat \rho}}_t( x ) $}

Let us consider the mapping analogous to Eq. \eqref{upsilonfp}
\begin{eqnarray}
  {\hat {\hat \rho}}_t( x )  = - \partial_x \left[ e^{ {\hat U}(x) }   {\hat \rho}_t( x ) \right]
\label{doubleupsilonfp}
\end{eqnarray}
The relation analogous to Eq. \eqref{fokkerplanckcurrentmap} using ${\hat U}(x) $ of Eq. \eqref{Uxhat}
\begin{eqnarray}
 {\hat J}_t( x )      = D(x) e^{- {\hat U}(x) }   {\hat {\hat \rho}}_t( x ) = e^{U(x) }  {\hat {\hat \rho}}_t( x )
\label{doublefokkerplanckcurrentmap}
\end{eqnarray}
yields that the boundary conditions of Eq. \eqref{BCdxJhat}
translate for the density ${\hat {\hat \rho}}_t( x ) $
\begin{eqnarray}
0 && = {\hat J}_t( x=0 ) =   e^{U(0) }   {\hat {\hat \rho}}_t( 0 )
 \nonumber \\
0 && = {\hat J}_t( x=L )  =   e^{U(L) }   {\hat {\hat \rho}}_t( L )
\label{doubleBCdxJhat}
\end{eqnarray}
into absorbing boundary conditions at $x=0$ and $x=L$.

The rule of Eq. \eqref{forcehat} yields that the force 
${ \hat{\hat F}}(x) $ actually coincide with the original force $F(x)$
\begin{eqnarray}
{ \hat{\hat F}}(x) = - {\hat F}(x) - D'(x) = F(x)
\label{doubleforcehat}
\end{eqnarray}
and that the corresponding potential ${\hat {\hat U}}(x) $ also coincide 
with the original potential $U(x)$ using Eq. \eqref{Uxhat}
 \begin{eqnarray} 
{\hat {\hat U}}(x) && =  - {\hat U}(x)  + \ln D(x) = U(x)
 \label{doubleUxhat}
\end{eqnarray}
so that the generator ${\hat {\hat {\cal L}}} $ coincide with the original generator ${\cal L} $
of Eq. \eqref{generator}
\begin{eqnarray}
{\hat {\hat {\cal L}}}  = {\cal L}
\label{doublegeneratorhat}
\end{eqnarray}
even if the boundary conditions are different.
Plugging Eq. \eqref{upsilonfp}
into Eq. \eqref{doubleupsilonfp} yields using ${\hat U}(x) $ of Eq. \eqref{Uxhat}
and ${\cal L} $
of Eq. \eqref{generator} 
\begin{eqnarray}
  {\hat {\hat \rho}}_t( x )  
&&  =  \partial_x \left[ e^{ {\hat U}(x) }  \partial_x \left( e^{ U(x) }   \rho_t( x ) \right)   \right]
  = \partial_x \left[ D(x) e^{ -U(x) } e^{U(x) }  \left(  U'(x) + \partial_x \right)   \rho_t( x )    \right]
  \nonumber \\
  && = \partial_x \left[ D(x)   \left(  U'(x) + \partial_x \right)   \rho_t( x )    \right] = {\cal L} \rho_t( x )
  = \partial_t \rho_t( x )
\label{doubleuhatrho}
\end{eqnarray}
 that the density ${\hat {\hat \rho}}_t( x )  $ actually
reduces to the time-derivative of the original density $\rho_t( x ) $.
In particular, it converges towards zero
\begin{eqnarray}
  {\hat {\hat \rho}}_*( x )  =0
\label{doubleuhatrho0}
\end{eqnarray}
as expected from the absorbing boundary conditions of Eq. \eqref{doubleBCdxJhat}.


\section{ Interpretation via supersymmetric quantum Hamiltonians }

\label{sec_susyFP}

The mapping from the boundary-driven non-equilibrium dynamics towards
some dual equilibrium dynamics \cite{tailleurMapping} that has been recalled in the previous section
 is reinterpreted in the present section
via the corresponding supersymmetric quantum Hamiltonians.

\subsection{ Supersymmetric quantum Hamiltonian $H =Q^{\dagger} Q$ associated with the non-equilibrium dynamics for $\rho_t(x) $ }

The change of variables involving the potential $U(x)$ of Eq. \eqref{Ux}
\begin{eqnarray}
\rho_t(x)  =  e^{ - \frac{ U(x)}{2}}  \psi_t(x )
\label{ppsi}
\end{eqnarray}
transforms the Fokker-Planck dynamics of Eq \ref{fokkerplanck} for the density $ \rho_t(x) $
 into the euclidean Schr\"odinger equation for $\psi_t(x )$
\begin{eqnarray}
-  \partial_t \psi_t(x)  = H \psi_t(x )
\label{schropsi}
\end{eqnarray}
where the quantum Hermitian Hamiltonian 
\begin{eqnarray}
 H = H^{\dagger} =  - \frac{ \partial  }{\partial x} D(x) \frac{ \partial  }{\partial x} +V(x)
\label{hamiltonien}
\end{eqnarray}
involves the scalar potential $V(x)$ 
\begin{eqnarray}
V(x) && \equiv D(x)  \frac{ [U'(x)]^2 }{4 }  -D(x) \frac{U''(x)}{2} -D'(x) \frac{U'(x)}{2}
\nonumber \\
&& = \frac{ F^2(x) }{4 D(x) } + \frac{F'(x)}{2}
\label{vfromu}
\end{eqnarray}
The Hamiltonian of Eq. \eqref{hamiltonien} can be factorized
into the well-known supersymmetric form 
(see the review \cite{review_susyquantum} and references therein)
\begin{eqnarray}
H =     Q^{\dagger} Q
\label{hsusy}
\end{eqnarray}
involving the first-order differential operator 
\begin{eqnarray}
Q   \equiv    \sqrt{ D(x) }  \left(\partial_x  +\frac{ U'(x)}{2 } \right)
\label{qsusy}
\end{eqnarray}
and its adjoint
\begin{eqnarray}
Q^{\dagger}  &&\equiv  \left(   - \partial_x  +\frac{ U'(x)}{2 } \right)\sqrt{ D(x) }
\label{adjoint}
\end{eqnarray}

The boundary conditions of Eq. \eqref{fokkerplanckreservoirs} for the density $\rho_t(x)$
translate via Eq. \eqref{ppsi} into the following boundary conditions for the wave-function $\psi_t(x)$
\begin{eqnarray}
  \psi_t( x=0 )   && =  c_0
  \nonumber \\
  \psi_t(x=L) && = e^{  \frac{ U(L)}{2}} c_L
\label{quantumreservoirs}
\end{eqnarray}

The discussion of Eqs \eqref{steadyeq} and \eqref{steadynoneq} 
can be thus translated as follows :

(i) when the steady current of Eq. \eqref{solusteadycurrent}
vanishes $J_*=0$, then the quantum ground state $\psi_*^{eq} (x) $ at zero energy
obtained from the equilibrium steady state $ \rho^{eq}_*(x) $
of Eq. \eqref{steadyeq}
via Eq. \eqref{ppsi}
\begin{eqnarray}
\psi_*^{eq} (x) = e^{  \frac{ U(x)}{2}}   \rho^{eq}_*(x) = c_0 e^{  - \frac{ U(x)}{2} } 
\ \ \ {\rm if } \ \  J_* =0
\label{GSsteadyeq}
\end{eqnarray}
is annihilated by the operator $Q$ of Eq. \eqref{qsusy}
\begin{eqnarray}
Q  \psi_*^{eq} (x)  =    \sqrt{ D(x) }  \left( \partial_x  +\frac{ U'(x)}{2 } \right)\psi_*^{eq} (x) =0
\label{qsusyannihilation}
\end{eqnarray}

(ii) when the steady current of Eq. \eqref{solusteadycurrent}
does not vanish $J_* \ne 0$, then the quantum ground state 
at zero energy
obtained from the non-equilibrium steady state $ \rho^{noneq}_*(x) $
of Eq. \eqref{steadynoneq}
via Eq. \eqref{ppsi}
\begin{eqnarray}
\psi_*^{noneq} (x) = e^{  \frac{ U(x)}{2}}   \rho^{noneq}_*(x) 
=  e^{ - \frac{ U(x)}{2}}   \frac{  c_0 \int_x^L dy \frac{ e^{U(y) }  } { D(y) } + c_Le^{U(L) } \int_0^x dy \frac{ e^{U(y) }  } { D(y) } }{ \int_0^L dz \frac{ e^{U(z) }  } { D(z) } } 
\label{GSsteadynoneq}
\end{eqnarray}
is not annihilated by the operator $Q$ of Eq. \eqref{qsusy}
\begin{eqnarray}
Q  \psi_*^{noneq} (x) && =    \sqrt{ D(x) }  \left( \partial_x  +\frac{ U'(x)}{2 } \right)\psi_*^{noneq} (x) 
=  \sqrt{ D(x) }e^{ - \frac{ U(x)}{2}}  \partial_x
 \frac{  c_0 \int_x^L dy \frac{ e^{U(y) }  } { D(y) } + c_Le^{U(L) } \int_0^x dy \frac{ e^{U(y) }  } { D(y) } }{ \int_0^L dz \frac{ e^{U(z) }  } { D(z) } } 
 \nonumber \\
 && = -
\left( \frac{   c_0   - c_Le^{U(L) }   }{ \int_0^L dz \frac{ e^{U(z) }  } { D(z) } } \right)
 \frac{e^{  \frac{ U(x)}{2}}  }{\sqrt{ D(x) } }
 \equiv - J_* \frac{e^{  \frac{ U(x)}{2}}  }{\sqrt{ D(x) } }
\label{qsusyNONannihilation}
\end{eqnarray}
where one recognizes the non-vanishing steady current $J_*$ of Eq. \eqref{solusteadycurrent}.
The state of Eq. \eqref{qsusyNONannihilation} is annihilated by the adjoint operator $Q^{\dagger}$
of Eq. \eqref{adjoint}
\begin{eqnarray}
Q^{\dagger} Q  \psi_*^{noneq} (x) && = - J_* Q^{\dagger} \frac{e^{  \frac{ U(x)}{2}}  }{\sqrt{ D(x) } }
=  - J_* \left(   - \partial_x  +\frac{ U'(x)}{2 } \right)\sqrt{ D(x) } \frac{e^{  \frac{ U(x)}{2}}  }{\sqrt{ D(x) } } =0
\label{qsusyNONannihilationadjoint}
\end{eqnarray}
as it should for $\psi_*^{noneq} (x) $ to be the zero-energy groundstate of the Hamiltonian of 
Eq. \eqref{hsusy}.


\subsection{ Supersymmetric quantum Hamiltonian ${\hat H} ={\hat Q}^{\dagger} {\hat Q}$ associated with the equilibrium dynamics for ${ \hat \rho}_t(x) $ }

Similarly, the change of variables 
analogous to Eq. \eqref{ppsi} for the new density ${\hat \rho}_t(x) $
that involves the new potential ${\hat U}(x)$ of Eq. \eqref{Uxhat}
\begin{eqnarray}
{\hat \rho}_t(x)  =  e^{ - \frac{ {\hat U}(x)}{2}}  {\hat \psi}_t(x )
\label{ppsihat}
\end{eqnarray}
transforms the Fokker-Planck dynamics of Eq. \eqref{fokkerplanckhat} for ${\hat \rho}_t(x)  $
into  into the euclidean Schr\"odinger equation for ${\hat \psi}_t(x )$
\begin{eqnarray}
-  \partial_t {\hat \psi}_t(x )  = {\hat H} {\hat \psi}_t(x )
\label{schropsihat}
\end{eqnarray}
where the quantum supersymmetric Hermitian Hamiltonian 
\begin{eqnarray}
 {\hat H} = {\hat H}^{\dagger} =  - \frac{ \partial  }{\partial x} D(x) \frac{ \partial  }{\partial x} +{\hat V}(x)
 = {\hat Q}^{\dagger} {\hat Q}
\label{hamiltonienhat}
\end{eqnarray}
involves the scalar potential associated with the force ${\hat F}(x) =  - F(x) - D'(x)$ of Eq. \eqref{forcehat}
\begin{eqnarray}
{\hat V}(x) && =  \frac{ {\hat F}^2(x) }{4 D(x) } + \frac{ {\hat F}'(x)  }{2} 
=  \frac{ F^2(x) }{4 D(x) } - \frac{F'(x)}{2} +\frac{F(x) D'(x) }{2D(x)}+ \frac{ [D'(x)]^2 }{4 D(x) }-\frac{D''(x)}{2}
\label{vfromuhat}
\end{eqnarray}
while the first-order differential operator ${\hat Q}$ associated with the potential ${\hat U}(x) $
of Eq. \eqref{Uxhat}
\begin{eqnarray}
{\hat Q}  \equiv    \sqrt{ D(x) }  \left( \partial_x  +\frac{ {\hat U}'(x)}{2 } \right)
=  \sqrt{ D(x) }  \left( \partial_x  +\frac{ \frac{ D'(x) }{D(x)} - U'(x)  }{2 } \right) 
= - \left(   - \partial_x  +\frac{ U'(x)}{2 } \right)\sqrt{ D(x) }
\equiv - Q^{\dagger}
\label{qsusyhat}
\end{eqnarray}
coincides with the opposite $(- Q^{\dagger})$ of the original adjoint operator $Q^{\dagger} $
of Eq. \eqref{adjoint}, so 
that the adjoint of Eq. \eqref{qsusyhat}
\begin{eqnarray}
{\hat Q}^{\dagger}  = \left(   - \partial_x  +\frac{ {\hat U}'}{2 } \right)\sqrt{ D(x) } = - Q
\label{qdaggersusyhat}
\end{eqnarray}
coincides with the opposite $(- Q)$ of the original operator $Q$
of Eq. \eqref{qsusy}.

The boundary conditions of Eq. \eqref{BChatrho}
for the density ${\hat \rho}_t(x)$
translate via Eq. \eqref{ppsihat} into the following boundary conditions for the wave-function ${\hat \psi}_t(x)$
\begin{eqnarray}
0 && 
=  D(0)  e^{ - \frac{ {\hat U}(0)}{2}} 
\left[ \frac{{\hat U}'(0)}{2}  {\hat \psi}_t( 0 )  + \partial_x {\hat \psi}_t( x ) \vert_{x=0} \right]
=  \sqrt{ D(0) } e^{ - \frac{ {\hat U}(0)}{2}} \left( {\hat Q} { \hat \psi}_t(x) \right) \vert_{x=0}
 \nonumber \\
0 && 
=    D(L)  e^{ - \frac{ {\hat U}(L)}{2}} 
\left[ \frac{{\hat U}'(L)}{2}  {\hat \psi}_t( L )  + \partial_x {\hat \psi}_t( x ) \vert_{x=L}\right]
=  \sqrt{ D(L) } e^{ - \frac{ {\hat U}(L)}{2}} \left( {\hat Q} { \hat \psi}_t(x) \right) \vert_{x=L}
\label{BChatpsi}
\end{eqnarray}

The equilibrium steady state ${\hat \rho}^{eq}_*( x ) $ of Eq. \eqref{hatsteady}
translate via Eq. \eqref{ppsihat}
into the quantum ground state ${\hat \psi}_*^{eq} (x) $ at zero energy
\begin{eqnarray}
{\hat \psi}_*^{eq} (x) =  e^{  \frac{ {\hat U}(x)}{2}} {\hat \rho}^{eq}_*( x ) 
= J_*  e^{ - \frac{ {\hat U}(x)}{2}} = J^* \frac{e^{  \frac{ U(x)}{2}}}{ \sqrt{D(x)} }
\label{psiGShat}
\end{eqnarray}
which is annihilated by the operator ${\hat Q} $
\begin{eqnarray}
{\hat Q}  {\hat \psi}_*^{eq} (x) = 0
\label{psiGShatannihilate}
\end{eqnarray}

Finally, the conservation of ${\hat N}  $ of Eq. \eqref{hatNt} reads for the quantum solution ${\hat \psi}_t(x ) $
\begin{eqnarray}
{\hat N} \equiv c_0 - e^{U(L)} c_L = \int_0^L  dx   e^{ - \frac{ {\hat U}(x)}{2}}  {\hat \psi}_t(x ) 
\label{hatNtpsi}
\end{eqnarray}


\subsection{ Relations between the two 
quantum supersymmetric Hamiltonians $H= Q^{\dagger} Q$ and ${\hat H}={\hat Q}^{\dagger} {\hat Q}$    }

Via the changes of variables of Eq. \eqref{ppsi}
and Eq. \eqref{ppsihat},
the correspondence of Eq. \eqref{upsilonfp}
between the two densities $\rho_t(x)$ and ${\hat \rho}_t(x ) $
translates into the following relation between the two quantum wavefunctions $\psi_t(x ) $ and ${\hat \psi}_t(x ) $
\begin{eqnarray}
{\hat \psi}_t(x ) =  e^{  \frac{ {\hat U}(x)}{2}}  {\hat \rho}_t(x) 
=  - e^{  \frac{\ln D(x) - U(x)  }{2}}  \partial_x \left[   e^{  \frac{ U(x)}{2}}  \psi_t(x ) \right]
= - \sqrt{ D(x) }  \left(\partial_x  +\frac{ U'(x)}{2 } \right)  \psi_t(x ) = - Q  \psi_t(x ) 
\label{psihatpsi}
\end{eqnarray}
while the reciprocal relation of Eq. \eqref{iniPfromhatP}
reads
\begin{eqnarray}
 c_0   -   e^{  \frac{ U(y)}{2}}  \psi_t(y ) =  \int_0^y dx  e^{ - \frac{ {\hat U}(x)}{2}}  {\hat \psi}_t(x )   
 =  \int_0^y dx   \frac{e^{  \frac{ U(x)}{2}}}{ \sqrt{D(x)} }  {\hat \psi}_t(x )   
\label{psipsihat}
\end{eqnarray}

For the corresponding zero-energy ground-states of
Eqs \eqref{GSsteadynoneq}
and \eqref{psiGShat}, 
 these relations become 
\begin{eqnarray}
{\hat \psi}_*^{eq}(x )   
= - \sqrt{ D(x) }  \left(\partial_x  +\frac{ U'(x)}{2 } \right)  \psi_*^{noneq}(x )  = - Q  \psi_*^{noneq}(x ) 
\label{psihatpsieq}
\end{eqnarray}
and 
\begin{eqnarray}
 c_0   -   e^{  \frac{ U(y)}{2}} \psi_*^{noneq}(y )  
 =  \int_0^y dx   \frac{e^{  \frac{ U(x)}{2}}}{ \sqrt{D(x)} }  {\hat \psi}_*^{eq}(x )   
\label{psipsihateq}
\end{eqnarray}

It is thus interesting to discuss the relations between the excited states
of the Hamiltonians $H$ and $\hat H$.


\subsubsection{ Properties of the excited states $\phi_n(x)$ of the 
Hamiltonian $H= Q^{\dagger} Q$     }

The excited states of $H$ are the eigenvectors $ \phi_n (x)$ associated with eigenvalues $E_n>0$
\begin{eqnarray}
E_n  \phi_n(x) = H    \phi_n(x) 
\label{hsusyeigenn}
\end{eqnarray}
satisfying the following absorbing boundary conditions (since the ground-state $\psi_*(x)$ of $H$ satisfies
the boundary conditions of Eq. \eqref{quantumreservoirs})
\begin{eqnarray}
\phi_n( 0 )  && =0
 \nonumber \\
\phi_n( L )  && =0
\label{BCexcitedpsi}
\end{eqnarray}
and the orthonormalization
\begin{eqnarray}
\delta_{n, n' } = \langle  \phi_n \vert  \phi_{n'} \rangle 
= \int_{0}^{L} dx \phi_n(x) \phi_{n'}(x) \ \ \ {\rm for } \ \ n=1,..+\infty 
\ \ {\rm and } \ \ \ n'=1,..,+\infty
\label{orthophinn}
\end{eqnarray}
These excited states are useful to write the spectral decomposition for the 
quantum propagator of the Hamiltonian $H$ in the presence of absorbing boundary conditions,
so it will be convenient to use the notation $H^{abs}$ to stress these absorbing boundary conditions
\begin{eqnarray}
  \langle x \vert  e^{-t H^{abs} } \vert x_0 \rangle
= \sum_{n=1}^{+\infty} e^{- t E_n} \phi_n(x)  \phi_n (x_0)
\label{HAbsspectral}
\end{eqnarray}


\subsubsection{ Properties of the excited states ${\hat \phi}_n(x)$ of the Hamiltonian ${\hat H}={\hat Q}^{\dagger} {\hat Q}$   }

 The excited states of $\hat H$ are the eigenvectors $ {\hat \phi}_n (x)$ associated  
to eigenvalues ${\hat E}_n>0$ 
\begin{eqnarray}
{\hat E}_n { \hat \phi}_n(x) = {\hat H}   { \hat \phi}_n(x) 
\label{hsusyeigen}
\end{eqnarray}
satisfying the boundary conditions of Eq. \eqref{BChatpsi}
\begin{eqnarray}
0 && 
=  D(0)  e^{ - \frac{ {\hat U}(0)}{2}} 
\left[ \frac{{\hat U}'(0)}{2}  { \hat \phi}_n( 0 )  + { \hat \phi}_n'( 0 )  \right]
= \sqrt{ D(0) } e^{ - \frac{ {\hat U}(0)}{2}} {\hat Q} { \hat \phi}_n(x) \vert_{x=0}
 \nonumber \\
0 && 
=    D(L)  e^{ - \frac{ {\hat U}(L)}{2}} 
\left[ \frac{{\hat U}'(L)}{2}  { \hat \phi}_n( L )  + { \hat \phi}_n'( x ) \right]
= \sqrt{ D(L) } e^{ - \frac{ {\hat U}(L)}{2}} {\hat Q} { \hat \phi}_n(x) \vert_{x=L}
\label{BChatpsin}
\end{eqnarray}
Meanwhile, the normalized ground state ${\hat \phi}_0(x) $ associated with zero-energy ${\hat E}_{n=0} $
involving the partition function $\hat Z $ of Eq. \eqref{zhat}
\begin{eqnarray}
  {\hat \phi}_0(x) && =   \frac{ e^{ - \frac{ {\hat U} (x) }{2} } }{\sqrt {\hat Z}} 
  =   \frac{ \frac{ e^{\frac{ U(x)}{2} }  } { \sqrt{ D(x) }} }{\sqrt {\hat Z}}
 \label{phi0hat}
\end{eqnarray}
is annihilated by $ {\hat Q} $.
Since the ground-state wave function $ {\hat \phi}_0(x) $ satisfies Eqs \eqref{hsusyeigen} and \eqref{BChatpsin} 
as all the excited states ${\hat \phi}_n(x) $ for $n>0$,
the orthonormalization relations involve the whole set of eigenstates $n=0,..+\infty $
\begin{eqnarray}
\delta_{n n' } = \langle { \hat \phi}_n \vert { \hat \phi}_{n'} \rangle 
= \int_{0}^{L} dx { \hat \phi}_n(x) { \hat \phi}_{n'}(x)
\ \ \ {\rm for } \ \ n=0,..+\infty 
\ \ {\rm and } \ \ \ n'=0,..,+\infty
\label{orthophin}
\end{eqnarray}
and the spectral decomposition for the 
quantum propagator of the Hamiltonian $\hat H$ reads
\begin{eqnarray}
  \langle x \vert  e^{-t {\hat H} } \vert x_0 \rangle
= \sum_{n=0}^{+\infty} e^{- t {\hat E}_n} \langle x \vert { \hat \phi}_n \rangle \langle { \hat \phi}_n \vert x_0 \rangle
\label{Hhatspectral}
\end{eqnarray}


\subsubsection{ Dual Hamiltonian ${\hat H}={\hat Q}^{\dagger} {\hat Q}=Q Q^{\dagger}$ as the
supersymmetric partner of the original Hamiltonian $H= Q^{\dagger} Q$  }

As a consequence of Eq. \eqref{qsusyhat},
 the supersymmetric partner $Q  Q^{\dagger} $ of the initial
Hamiltonian $H= Q^{\dagger} Q$ coincides with the dual Hamiltonian ${\hat H}={\hat Q}^{\dagger} {\hat Q}$
\begin{eqnarray}
Q  Q^{\dagger} = {\hat Q}^{\dagger} {\hat Q} = {\hat H} 
\label{hsusypartnerhatdouble}
\end{eqnarray}
The standard procedure described in the review \cite{review_susyquantum} and references
therein, where the excited states ${\hat \phi}_n (x)$ of $ {\hat H} $ for $n>0$ are constructed from the
corresponding excited states $ \phi_n (x)$ of $  H $ via the application of the operator $Q$
\begin{eqnarray}
{\hat \phi}_n (x) && =  \frac{ Q \phi_n(x)   }{ \sqrt{E_n} }
\nonumber \\
{\hat E}_n && = E_n
\label{phinviaQ}
\end{eqnarray}
is clearly compatible with the boundary conditions of Eqs \eqref{BCexcitedpsi}
and \eqref{BChatpsin}.
Indeed, if $ \phi_n (x)$ satisfies the eigenvalue Eq. \eqref{hsusyeigenn}
with the boundary conditions of Eq. \eqref{BCexcitedpsi} and is normalized 
$\langle \phi_n \vert \phi_n \rangle =1$,
then the state ${\hat \phi}_n (x)$ of Eq. \eqref{phinviaQ}
is an eigenstate of ${\hat H}={\hat Q}^{\dagger}{\hat Q}=Q Q^{\dagger} $ for the same energy ${\hat E}_n=E_n$
 \begin{eqnarray}
{\hat H}{\hat \phi}_n (x)  && =Q Q^{\dagger}  \frac{ Q \phi_n(x)   }{ \sqrt{E_n} }
= Q   \frac{ H \phi_n(x)   }{ \sqrt{E_n} } = E_n  \frac{ Q \phi_n(x)   }{ \sqrt{E_n} }
= E_n {\hat \phi}_n (x)
\label{phinviaQeigenhat}
\end{eqnarray}
that
satisfies the boundary conditions of Eq. \eqref{BChatpsin} 
using $ {\hat Q}=-Q^{\dagger} $ of Eq. \eqref{qsusyhat}
\begin{eqnarray}
 {\hat Q} {\hat \phi}_n (x) && =  \frac{ - Q^{\dagger} Q \phi_n(x)   }{ \sqrt{E_n} } = 
 \frac{ - H \phi_n(x)   }{ \sqrt{E_n} } = - \sqrt{E_n} \phi_n(x)
\label{phinviaQhat}
\end{eqnarray}
while it is normalized
\begin{eqnarray}
\langle {\hat \phi} \vert {\hat \phi} \rangle  && 
=  \frac{ \langle \phi_n \vert Q^{\dagger}Q \vert \phi_n \rangle }{E_n} 
=  \frac{ \langle \phi_n \vert H \vert \phi_n \rangle }{E_n} 
 = \frac{E_n}{E_n}=1
\label{phinviaQnormahat}
\end{eqnarray}

The relation of Eq. \eqref{phinviaQ} can be inverted by 
solving the corresponding differential equation for $\phi_n(x)$ 
involving the operator $Q$ of Eq. \eqref{qsusy}
\begin{eqnarray}
\sqrt{E_n} {\hat \phi}_n (x)  =   Q \phi_n(x) 
= \sqrt{ D(x) }  \left(\partial_x  +\frac{ U'(x)}{2 } \right) \phi_n(x)
= \sqrt{ D(x) }  e^{ - \frac{ U(x)}{2}} \partial_x  \left(  e^{  \frac{ U(x)}{2}}  \phi_n(x) \right)
\label{phinviaQdiff}
\end{eqnarray}
The integration using the boundary condition $ \phi_n(x=0)=0 $ of Eq. \eqref{BCexcitedpsi}
yields the solution 
\begin{eqnarray}
\phi_n(y) =
\sqrt{E_n} e^{ - \frac{ U(y)}{2}}   \int_0^y dx {\hat \phi}_n (x)  \frac{e^{  \frac{ U(x)}{2}}}{\sqrt{ D(x) }} 
=    \sqrt{E_n} e^{ - \frac{ U(y)}{2}}   \int_0^y dx {\hat \phi}_n (x)  e^{ - \frac{ {\hat U}(x)}{2}}
\label{phinviaQdiffsolu}
\end{eqnarray}
The last rewriting in terms of the dual potential $ {\hat U}(x)$ of Eq. \eqref{Uxhat}
is useful to check the vanishing of $\phi_n(y=L) =0$
since the integral for $y=L$ vanishes as a consequence of the orthogonality of 
any excited state ${\hat \phi}_{n>0}(x) $ with the ground-state 
${\hat \phi}_{n=0} (x)$ of Eq. \eqref{phi0hat}.


\subsubsection{ Original Hamiltonian $H= Q^{\dagger} Q= {\hat Q} {\hat Q}^{\dagger}$
 as the
supersymmetric partner of the dual Hamiltonian ${\hat H}={\hat Q}^{\dagger} {\hat Q}$  }

As a consequence of Eq. \eqref{qsusyhat},
 the supersymmetric partner ${\hat Q} {\hat Q}^{\dagger} $ of the 
dual Hamiltonian ${\hat H}={\hat Q}^{\dagger} {\hat Q}$
coincides with the original Hamiltonian
$H= Q^{\dagger} Q$  
\begin{eqnarray}
{\hat Q} {\hat Q}^{\dagger}= Q^{\dagger} Q= H
\label{hsusypartnerhat}
\end{eqnarray}
The boundary conditions of Eqs \eqref{BCexcitedpsi}
and \eqref{BChatpsin} are again compatible with the standard procedure to
construct the excited states 
(see the review \cite{review_susyquantum} and references therein) that corresponds to the relation already written in Eq. \eqref{phinviaQhat} 
\begin{eqnarray}
 \phi_n (x) && = -  \frac{ {\hat Q} {\hat \phi}_n (x)  }{ \sqrt{E_n} } 
\label{phinviaQhatbis}
\end{eqnarray}


\subsection{ Simplest example : diffusion coefficient $D(x)=1$ and constant force $F(x) =  f$ on the interval $]0, L[$  }

\label{subsec_constant}

Let us now illustrate the general formalism described above
for the simplest example of the Fokker-Planck dynamics involving
the constant diffusion coefficient $D(x)=1$ and the constant force $F(x) =  f$ on the whole interval $]0, L[$.

\subsubsection{ Excited states of the original Hamiltonian $H$  }

For the diffusion coefficient $D(x)=1$
and the constant force $ F(x)=f$, the potential $U(x)$ of Eq. \eqref{Ux} is linear
\begin{eqnarray}
U(x)  = - f x
\label{Uxlinear}
\end{eqnarray}
while the quantum potential of Eq. \eqref{vfromu} reduces to the constant
\begin{eqnarray}
V(x)  = \frac{ F^2(x) }{4  } + \frac{F'(x)}{2} = \frac{ f^2 }{4  } 
\label{vfromuxlinear}
\end{eqnarray}
and the first-order operators of Eqs \eqref{qsusy} and \eqref{adjoint} read
\begin{eqnarray}
Q   && \equiv    \sqrt{ D(x) }  \left(\partial_x  +\frac{ U'(x)}{2 } \right) = \partial_x  - \frac{ f }{2 }
\nonumber \\
Q^{\dagger}  &&\equiv  \left(   - \partial_x  +\frac{ U'(x)}{2 } \right)\sqrt{ D(x) } = - \partial_x  - \frac{ f }{2 }
\label{qsusyadjoint}
\end{eqnarray}

The eigenvalue equation for the quantum Hamiltonian $H$ of Eq. \eqref{hamiltonien} 
with the potential of Eq. \eqref{vfromuxlinear}
\begin{eqnarray}
E  \phi(x) = H \phi(x)  =  - \phi'' (x) + \frac{ f^2 }{4  }  \phi(x)
\label{Ephi}
\end{eqnarray}
and with the absorbing boundary conditions of Eq. \eqref{BCexcitedpsi} at $x=0$ and at $x=L$ 
\begin{eqnarray}
0  && =  \phi(x=0) 
\nonumber \\
0 && =   \phi(x=L)
\label{absorbting0l}
\end{eqnarray}
leads to the orthonormalized eigenstates
\begin{eqnarray}
\phi_n(x)  = \sqrt{\frac{2}{L} }  \sin\left( n \frac{ \pi }{L} x \right) \ \ \ {\rm with } \ \  n=1,2,..,+\infty
\label{phinabs}
\end{eqnarray}
associated with the eigenvalues
\begin{eqnarray}
E_n= \frac{f^2 }{4  } + n^2 \frac{ \pi^2 }{L^2}\ \ \ {\rm with } \ \  n=1,2,..,+\infty
\label{Enabs}
\end{eqnarray}


\subsubsection{ Excited states of the dual Hamiltonian $\hat H$  }

For the diffusion coefficient $D(x)=1$
and the constant force $ F(x)=f$, the new force $ {\hat F}(x)$ of Eq. \eqref{forcehat}
reduces to the constant
\begin{eqnarray}
{\hat F}(x)= - F(x)  = - f
\label{forcehatconstant}
\end{eqnarray}
with the corresponding linear potential $  {\hat U}(x) $ of Eq. \eqref{Uxhat}
 \begin{eqnarray}
{\hat U}(x) && =  - U(x)  = f x
 \label{Uxhatlin}
\end{eqnarray}
The quantum potential of Eq. \eqref{vfromu} is constant and coincides with Eq. \eqref{vfromuxlinear}
\begin{eqnarray}
{\hat V}(x)  = \frac{ {\hat F}^2(x) }{4  } + \frac{{\hat F}'(x)}{2} = \frac{ f^2 }{4  } = V(x)
\label{vfromuxlinearhat}
\end{eqnarray}
while the first-order operators of Eqs \eqref{qsusyhat} and \eqref{qdaggersusyhat} read
\begin{eqnarray}
{\hat Q}  && =    \sqrt{ D(x) }  \left( \partial_x  +\frac{ {\hat U}'(x)}{2 } \right)
= \partial_x  + \frac{f}{2} =  - Q^{\dagger}
\nonumber \\
{\hat Q}^{\dagger} && = \left(   - \partial_x  +\frac{ {\hat U}'}{2 } \right)\sqrt{ D(x) } 
= - \partial_x  + \frac{f}{2} 
= - Q
\label{qqdaggersusyhat}
\end{eqnarray}

The eigenvalue equation for the quantum Hamiltonian ${\hat H}$ 
with the potential of Eq. \eqref{vfromuxlinearhat}
\begin{eqnarray}
{\hat E}  {\hat \phi}(x) = {\hat H}  {\hat \phi}(x)  =  -  {\hat \phi}''(x) + \frac{ f^2 }{4  }  {\hat \phi}(x)
\label{Ephihat}
\end{eqnarray}
and with the boundary conditions of Eqs \eqref{BChatpsin}
at $x=0$ and at $x=L$ 
\begin{eqnarray}
0  && =     \frac{f}{2} {\hat \phi}(0) + {\hat \phi}'(0)
\nonumber \\
0 && =   \frac{f}{2} {\hat \phi}(L) + {\hat \phi}'(L) 
\label{reflecting0l}
\end{eqnarray}
leads to the orthonormalized eigenstates
\begin{eqnarray}
{\hat \phi}_n(x)  = \sqrt{\frac{2}{L} {\hat E_n}} 
\left[ n \frac{ \pi }{L} \cos\left( n \frac{ \pi }{L} x \right) - \frac{f}{2} \sin\left( n \frac{ \pi }{L} x \right) \right] \ \ \ {\rm with } \ \  n=1,2,..,+\infty
\label{phinabshat}
\end{eqnarray}
associated with the eigenvalues that coincide with the eigenvalues $E_n$ of Eq. \eqref{Enabs}
\begin{eqnarray}
{\hat E_n}= \frac{f^2 }{4  } + n^2 \frac{ \pi^2 }{L^2}=E_n \ \ \ {\rm with } \ \  n=1,2,..,+\infty
\label{Enabshat}
\end{eqnarray}

One can check that the eigenstates of Eqs \eqref{phinabs} and \eqref{phinabshat}
satisfy Eqs \eqref{phinviaQ} and \eqref{phinviaQhatbis} as it should
\begin{eqnarray}
\sqrt{E_n} {\hat \phi}_n (x) && = Q \phi_n(x)  = \left( \partial_x  - \frac{ f }{2 } \right) \phi_n(x)
\nonumber \\
\sqrt{E_n} \phi_n (x) && = -   {\hat Q} {\hat \phi}_n (x)  = - \left( \partial_x  + \frac{ f }{2 } \right){\hat \phi}_n (x)
\label{phinviaQcheck}
\end{eqnarray}


\section{  Spectral decompositions of the Fokker-Planck dynamics }

\label{sec_spectralFP}

In this section, we use the spectral properties of the quantum 
supersymmetric quantum Hamiltonians to analyze the spectral decompositions of the
corresponding Fokker-Planck dynamics.

\subsection{ Spectral decomposition for the quantum solution ${\hat \psi}_t( x )$  associated with  
the dual Hamiltonian ${\hat H}={\hat Q}^{\dagger} {\hat Q}$  }

The spectral decomposition of the quantum propagator $ \langle x \vert e^{- t {\hat H} } \vert x_0 \rangle $  of Eq. \eqref{Hhatspectral}
is useful to write the solution ${\hat \psi}_t(x) $ at time $t$ 
 in terms of the initial condition ${\hat \psi}_{t=0}( x )$ 
\begin{eqnarray}
 {\hat \psi}_t( x ) && = \int_0^L dx_0 \langle x \vert e^{- t {\hat H} } \vert x_0 \rangle  {\hat \psi}_0( x_0 )
 =   \sum_{n=0}^{+\infty} e^{- t {\hat E}_n} { \hat \phi}_n (x) 
\int_0^L dx_0  { \hat \phi}_n (x_0 )    {\hat \psi}_0( x_0 )
\label{Convolpsihatspectral}
\end{eqnarray}
The convergence towards ${\hat \psi}^{eq}_*( x ) $ of Eq. \eqref{psiGShat}
 for $t \to + \infty$ can be recovered 
 using the constant $ {\hat N} $ of Eq. \eqref{hatNtpsi} for the initial condition ${\hat \psi}_{t=0}( x_0 ) $
\begin{eqnarray}
 {\hat \psi}_t( x ) && \opsimeq_{t \to + \infty}    { \hat \phi}_0 (x) 
\int_0^L dx_0  { \hat \phi}_0 (x_0 )    {\hat \psi}_0( x_0 )
=   \frac{ e^{ - \frac{ {\hat U} (x) }{2} } }{\hat Z} 
\int_0^L dx_0  e^{ - \frac{ {\hat U} (x_0) }{2} }   {\hat \psi}_0( x_0 )
\nonumber \\
&& =  \frac{ e^{ - \frac{ {\hat U} (x) }{2} } }{\hat Z} {\hat N}
=  {\hat N} \frac{ \frac{ e^{\frac{ U(x)}{2} }  } { \sqrt{ D(x) }} }{\int_0^L dy \frac{ e^{U(y) }  } { D(y) }}   \equiv {\hat \psi}^{eq}_*( x )
\label{Convolpsihateq}
\end{eqnarray}

In summary, the convergence of $ {\hat \psi}_t( x ) $ towards ${\hat \psi}^{eq}_*( x ) $
\begin{eqnarray}
 {\hat \psi}_t( x ) && 
 ={\hat \psi}^{eq}_*( x ) +   \sum_{n=1}^{+\infty} e^{- t {\hat E}_n} { \hat \phi}_n (x) 
\int_0^L dx_0  { \hat \phi}_n (x_0 )    {\hat \psi}_0( x_0 )
\label{psihatspectral}
\end{eqnarray}
involves the energies ${\hat E}_n >0$ and the corresponding excited states $ { \hat \phi}_n (x) $.


\subsection{ Spectral decomposition of the density 
${\hat \rho}_t(x) $ relaxing towards the equilibrium steady state ${\hat \rho}_*^{eq}(x) $  }

The translation of Eq. \eqref{psihatspectral}
for the density ${\hat \rho}_t(x) $ via Eq. \eqref{ppsihat} reads
\begin{eqnarray}
{\hat \rho}_t(x)  =  e^{ - \frac{ {\hat U}(x)}{2}} {\hat \psi}_t( x ) && 
 ={\hat \rho}^{eq}_*( x ) +   \sum_{n=1}^{+\infty} e^{- t {\hat E}_n} \left[ e^{ - \frac{ {\hat U}(x)}{2}} { \hat \phi}_n (x) \right]
\int_0^L dx_0 \left[ { \hat \phi}_n (x_0 )    e^{  \frac{ {\hat U}(x_0)}{2}} \right] {\hat \rho}_0( x_0 )
\nonumber \\
&&  ={\hat \rho}^{eq}_*( x ) +   \sum_{n=1}^{+\infty} 
e^{- t {\hat E}_n} \left[  {\hat \phi}_0(x){ \hat \phi}_n (x)  \right]
\int_0^L dx_0 \left[ \frac{{ \hat \phi}_n (x_0 )}{ {\hat \phi}_0(x_0)}  \right] {\hat \rho}_0( x_0 )
\label{rhohatspectral}
\end{eqnarray}

The interpretation in terms of propagator $\langle x \vert e^{t {\hat {\cal L}} } \vert x_0 \rangle $ associated with the generator
$ {\hat {\cal L}}$
\begin{eqnarray}
 {\hat \rho}_t( x ) = \int_0^L dx_0 \langle x \vert e^{t {\hat {\cal L}} } \vert x_0 \rangle  {\hat \rho}_0( x_0 )
\label{Convolhat}
\end{eqnarray}
yields that the eigenvalues of the generator ${\hat {\cal L}} $ are the opposite $[- {\hat E}_n] $
of the energies ${\hat E}_n$, 
while
\begin{eqnarray}
{\hat r}_n(x) && \equiv {\hat \phi}_0(x){ \hat \phi}_n (x) 
=  \frac{ e^{ - \frac{ {\hat U} (x) }{2} } }{\sqrt {\hat Z}} { \hat \phi}_n (x) 
\nonumber \\
 {\hat l}_n ( x_0 )  && \equiv \frac{{ \hat \phi}_n (x_0 )}{ {\hat \phi}_0(x_0)}
 = \sqrt {\hat Z} e^{  \frac{ {\hat U} (x_0) }{2} } { \hat \phi}_n (x_0 )
\label{rnlnphin}
\end{eqnarray}
are the corresponding right and left eigenvectors satisfying the eigenvalue equations
\begin{eqnarray}
 -{\hat E}_n {\hat r}_n(x) &&={\hat {\cal L}} {\hat r}_n(x) 
 \nonumber \\
  -{\hat E}_n {\hat l}_n(x) && = {\hat {\cal L}}^{\dagger} {\hat l}_n(x)   
 \label{fokkerplanckeigen}
\end{eqnarray}
The boundary conditions of Eq. \eqref{BChatpsin} 
can be translated for the right eigenvectors
\begin{eqnarray}
0 && = D(0) \left[ {\hat U}'(0)  { \hat r}_n( 0 )  + { \hat r}_n'( 0 )  \right]  
 \nonumber \\
  0 && = D(L) \left[ {\hat U}'(L)  { \hat r}_n( L )  + { \hat r}_n'( L )  \right] 
\label{BChatrn}
\end{eqnarray}
and for the left eigenvectors
\begin{eqnarray}
0 && =  D(0)  e^{ -  {\hat U}(0)}  { \hat l}_n'( 0 )
 \nonumber \\
 0 && =     D(L)  e^{ -  {\hat U}(L)}  { \hat l}_n'( x ) 
\label{BChatln}
\end{eqnarray}
while the orthonormalization translated from Eq. \eqref{orthophin} reads
\begin{eqnarray}
\delta_{n n' } = \langle { \hat \phi}_n \vert { \hat \phi}_{n'} \rangle 
=  \langle { \hat l}_n \vert { \hat r}_{n'} \rangle
\ \ \ {\rm for } \ \ n=0,..+\infty 
\ \ {\rm and } \ \ \ n'=0,..,+\infty
\label{orthorln}
\end{eqnarray}

For the vanishing eigenvalue ${\hat E}_0=0$,  the right and left eigenvectors reduce to
\begin{eqnarray}
  {\hat r}_0(x) &&= {\hat \phi}^2_0(x)= \frac{ e^{ -  {\hat U} (x) } }{\hat Z} = 
  \frac{ \frac{ e^{ U(x) }  } {  D(x) } }{ \int_0^L dy \frac{ e^{U(y) }  } { D(y) }}
 \nonumber \\
  {\hat l}_0(x_0) && =    1
 \label{r0l0}
\end{eqnarray}

In summary, the spectral decomposition of Eq. \eqref{rhohatspectral} can be rewritten as
\begin{eqnarray}
 {\hat \rho}_t( x ) 
 && = {\hat \rho}^{eq}_*( x ) +
 \sum_{n=1}^{+\infty} e^{- t {\hat E}_n} {\hat r}_n(x)
 \int_0^L dx_0   {\hat l}_n(x_0)   {\hat \rho}_0( x_0 )
\label{Convolhatspectralsteady}
\end{eqnarray}


\subsection{ Spectral decomposition for the quantum solution $ \psi_t( x )$  associated with  
the original Hamiltonian $H=Q^{\dagger}Q$  }


\subsubsection{ Spectral decomposition for $\psi_t(y )  $ 
using the propagator $\langle x \vert  e^{-t H^{abs} } \vert x_0 \rangle $ in the presence of absorbing boundary conditions   }

As explained around Eq. \eqref{HAbsspectral}, the excited states $\phi_n(x)$ of $H$
are involved in the spectral decomposition of the propagator 
$\langle x \vert  e^{-t H^{abs} } \vert x_0 \rangle$ associated with $H$
in the presence of absorbing boundary conditions.
It is thus convenient to introduce the difference $\psi^{abs}_t(x) $ between $\psi_t(x) $ and the non-equilibrium steady state
$\psi_*^{noneq}(x) $ of Eq. \eqref{GSsteadynoneq}
\begin{eqnarray}
\psi_t^{abs}(x) \equiv  \psi_t(x) - \psi_*^{noneq}(x)
\label{psiabs}
\end{eqnarray}
Its spectral decomposition involving the propagator of Eq. \eqref{HAbsspectral}
\begin{eqnarray}
 \psi_t^{abs} ( x ) && 
 = \int_0^L dx_0 \langle x \vert e^{- t  H^{abs} } \vert x_0 \rangle   \psi^{abs}_0( x_0 )
 \nonumber \\
&&  =  \sum_{n=1}^{+\infty} e^{- t E_n} \phi_n(x) \int_0^L dx_0   \phi_n (x_0)  \psi^{abs}_0( x_0 )
\label{psibsspectral}
\end{eqnarray}
yields for the full solution $\psi_t(x)$ in terms of its initial condition $ \psi_{t=0}(x) $
\begin{eqnarray}
  \psi_t(x) && = \psi_*^{noneq}(x) + \psi_t^{abs}(x)
\nonumber \\
&&   = \psi_*^{noneq}(x) 
  + \sum_{n=1}^{+\infty} e^{- t E_n} \phi_n(x) \int_0^L dx_0   \phi_n (x_0) 
  \left[ \psi_0(x_0) - \psi_*^{noneq}(x_0) \right]
\label{psitspectral}
\end{eqnarray}

In the following subsection, it is interesting to check the consistency
between the spectral decomposition for $ \psi_t^{abs} ( x ) $ of Eq. \eqref{psibsspectral}
and of ${\hat \psi}_t( x ) $ Eq. \eqref{psihatspectral} via the supersymmetric-partnership
relations between their excited eigenvectors discussed in the previous section.


\subsubsection{ Spectral decomposition for $\psi^{abs}_t(y )  $ 
from the spectral decomposition for ${\hat \psi}_t( x ) $   }

The relation of Eq. \eqref{psipsihat} between 
$ \psi_t( x )$ and ${\hat \psi}_t( x ) $ reads for the difference of Eq. \eqref{psiabs}
\begin{eqnarray}
\psi_t^{abs}(y) =   - \psi_*^{noneq}(y) + \psi_t(y)
=  - \psi_*^{noneq}(y) 
+e^{  - \frac{ U(y)}{2}}  \left[ c_0 - \int_0^y dx  e^{ - \frac{ {\hat U}(x)}{2}}  {\hat \psi}_t(x )    \right]
\label{psiabshat}
\end{eqnarray}
so that one can plug the spectral decomposition for ${\hat \psi}_t(x )  $ of Eq. \eqref{psihatspectral}
to obtain
\begin{eqnarray}
\psi_t^{abs}(y) && =   - \psi_*^{noneq}(y) + \psi_t(y)
=  - \sum_{n=1}^{+\infty} e^{- t {\hat E}_n} e^{  - \frac{ U(y)}{2}}  \int_0^y dx  e^{ - \frac{ {\hat U}(x)}{2}} 
  { \hat \phi}_n (x) 
\int_0^L dx_0  { \hat \phi}_n (x_0 )    {\hat \psi}_0( x_0 )
\nonumber \\
&& = - \sum_{n=1}^{+\infty} e^{- t  E_n} \frac{\phi_n(y)}{ \sqrt{E_n} }
\int_0^L dx_0  { \hat \phi}_n (x_0 )    {\hat \psi}_0( x_0 )
\label{psiabshatspec}
\end{eqnarray}
where we have recognized the eigenvector $\phi_n(y) $ of Eq. \eqref{phinviaQdiffsolu}.
We now need to use the relation of Eq. \eqref{psihatpsi} to rewrite the initial condition
${\hat \psi}_0(x )  $ in terms of the initial condition $\psi_0(x )  $ and then $\psi_0(x )  $
via Eq. \eqref{psiabs}
\begin{eqnarray}
{\hat \psi}_0(x )  = - Q  \psi_0(x ) = -Q \left( \psi_*^{noneq}(x )+\psi^{abs}_0(x )\right)
= {\hat \psi}_*^{eq}(x ) - \sqrt{ D(x) }  \left(\partial_x  +\frac{ U'(x)}{2 } \right)  \psi^{abs}_0(x )  
\label{psihatpsi0}
\end{eqnarray}
Since $ {\hat \psi}_*^{eq}(x ) $ of Eq. \eqref{psiGShat}
is proportional to $ { \hat \phi}_{n=0} (x_0 ) $ of Eq. \eqref{phi0hat}
that is orthogonal to any excited state $ { \hat \phi}_n (x ) $,
the last integral of Eq. \eqref{psiabshatspec} reduces to the contribution involving $\psi^{abs}_0(x_0 ) $
that can be rewritten via integration by parts using the vanishing of $ \psi^{abs}_0(x_0 )$ at the two boundaries as
\begin{eqnarray}
&& \int_0^L dx_0  { \hat \phi}_n (x_0 )    {\hat \psi}_0( x_0 )
 = - \int_0^L dx_0  { \hat \phi}_n (x_0 )  \sqrt{ D(x_0) }  \left(\partial_{x_0}  +\frac{ U'(x_0)}{2 } \right)  \psi^{abs}_0(x_0 )  
\nonumber \\
&& = - \int_0^L dx_0  \psi^{abs}_0(x_0 ) 
   \left( - \partial_{x_0}  +\frac{ U'(x_0)}{2 } \right) \sqrt{ D(x_0) }  { \hat \phi}_n (x_0 )  
   = - \int_0^L dx_0  \psi^{abs}_0(x_0 ) Q^{\dagger}_{x_0}  { \hat \phi}_n (x_0 )  
\label{integparts}
\end{eqnarray}
where we have recognized the adjoint operator $Q^{\dagger}_{x_0} $ of Eq. \eqref{adjoint}.
So one can use Eq. \eqref{phinviaQ} to replace $ { \hat \phi}_n (x_0 ) $
in terms of $ \phi_n(x)$ to obtain
\begin{eqnarray}
 Q^{\dagger}_{x_0}  { \hat \phi}_n (x_0 )  = Q^{\dagger}_{x_0} \frac{ Q_{x_0} \phi_n(x_0)   }{ \sqrt{E_n} } = \frac{ H_{x_0} \phi_n(x_0)   }{ \sqrt{E_n} } = \sqrt{E_n}\phi_n(x_0)
\label{integpartsinside}
\end{eqnarray}
Putting everything together, Eq. \eqref{psiabshatspec} becomes
\begin{eqnarray}
\psi_t^{abs}(y) &&  =  \sum_{n=1}^{+\infty} e^{- t  E_n} \phi_n(y)
\int_0^L dx_0  \psi^{abs}_0(x_0 )\phi_n(x_0)
\label{psiabshatspecfinal}
\end{eqnarray}
in agreement with Eq. \eqref{psibsspectral}.


\subsection{ Spectral decomposition for the density 
$ \rho_t(x) $ relaxing towards the non-equilibrium steady state $ \rho_*^{noneq}(x) $  }

The spectral decomposition of $\psi_t(x ) $ Eq. \eqref{psitspectral}
can be translated
for the density $ \rho_t(x) $ via Eq. \eqref{ppsi} and Eq. \eqref{GSsteadynoneq}
\begin{eqnarray}
\rho_t(x)  && =  e^{ - \frac{ U(x)}{2}}  \psi_t(x )
= e^{ - \frac{ U(x)}{2}} \psi_*^{noneq}(x) 
  + \sum_{n=1}^{+\infty} e^{- t E_n} e^{ - \frac{ U(x)}{2}} \phi_n(x) \int_0^L dx_0   \phi_n (x_0) 
  \left[ \psi_0(x_0) - \psi_*^{noneq}(x_0) \right]
\nonumber \\
&& =\rho^{noneq}_*(x) 
 + \sum_{n=1}^{+\infty} e^{- t E_n} e^{ - \frac{ U(x)}{2}} \phi_n(x) \int_0^L dx_0   \phi_n (x_0) 
 e^{  \frac{ U(x_0)}{2}} \left[ \rho_0(x_0) - \rho_*^{noneq}(x_0) \right]
\label{rhotspectral}
\end{eqnarray}

The physical interpretation is that the spectral decomposition is clearer for the difference
analogous to Eq. \eqref{psiabs}
\begin{eqnarray}
\rho_t^{abs}(x) && \equiv  \rho_t(x) - \rho_*^{noneq}(x)
\nonumber \\
&& = \sum_{n=1}^{+\infty} e^{- t E_n} e^{ - \frac{ U(x)}{2}} \phi_n(x) \int_0^L dx_0   \phi_n (x_0) 
 e^{  \frac{ U(x_0)}{2}}  \rho^{abs}_0(x) 
 \equiv \int_0^L dx_0 \langle x \vert e^{ t  {\cal L}^{abs} } \vert x_0 \rangle  \rho^{abs}_0(x)  
\label{rhoabs}
\end{eqnarray}
that involves the propagator $\langle x \vert e^{ t  {\cal L}^{abs} } \vert x_0 \rangle $
associated with
Fokker-Planck generator ${\cal L}^{abs}$ in the presence of absorbing boundary conditions
\begin{eqnarray}
\langle x \vert e^{ t  {\cal L}^{abs} } \vert x_0 \rangle 
= \sum_{n=1}^{+\infty} e^{- t E_n} e^{ - \frac{ U(x)}{2}} \phi_n(x) 
 e^{  \frac{ U(x_0)}{2}}  \phi_n (x_0 ) 
 \equiv \sum_{n=1}^{+\infty} e^{- t E_n} r_n(x)    l_n (x_0) 
\label{calLabsspectral}
\end{eqnarray}
where, up to some normalization $K_n$ that will be chosen later, one can interpret 
\begin{eqnarray}
r_n(x) && \equiv K_n e^{ - \frac{ U(x)}{2}} \phi_n(x) 
\nonumber \\
l_n ( x_0 )  && \equiv \frac{1}{K_n} e^{  \frac{ U(x_0)}{2}}  \phi_n (x_0 ) 
\label{rnlnphinans}
\end{eqnarray}
as the corresponding right and left eigenvectors satisfying the eigenvalue equations
\begin{eqnarray}
 -E_n r_n(x) &&= {\cal L} r_n(x) 
 \nonumber \\
  -E_n l_n(x) && = {\cal L}^{\dagger} l_n(x)   
 \label{fokkerplanckeigenabs}
\end{eqnarray}
and the absorbing boundary conditions inherited from Eq. \eqref{BCexcitedpsi}
\begin{eqnarray}
r_n( 0 )  && =0 = r_n(L)
 \nonumber \\
l_n( 0 )  && =0 = l_n(L)
\label{BCexcitedpsirl}
\end{eqnarray}
while the orthonormalization translated from Eq. \eqref{orthophinn} reads
\begin{eqnarray}
\delta_{n, n' } = \langle  \phi_n \vert  \phi_{n'} \rangle 
= \langle  l_n \vert  r_{n'} \rangle  \ \ \ {\rm for } \ \ n=1,..+\infty 
\ \ {\rm and } \ \ \ n'=1,..,+\infty
\label{orthophinnrl}
\end{eqnarray}

The relations of Eqs \eqref{phinviaQ} and \eqref{phinviaQhatbis}
between the eigenstates $\phi_n(x)$ and $ {\hat \phi}_n(x)$
lead to the following relations between the right and left eigenvectors of
Eqs \eqref{rnlnphin} and \eqref{rnlnphinans} :

(i) the right eigenvector ${\hat r}_n(x)  $ written in terms of the right eigenvector $r_n(x)$
\begin{eqnarray}
{\hat r}_n(x) && 
=  \frac{ e^{ - \frac{ {\hat U} (x) }{2} } }{\sqrt {\hat Z}} { \hat \phi}_n (x) 
= \frac{ e^{  \frac{ U (x) }{2} } }{\sqrt { E_n {\hat Z} D(x) }}  
 \sqrt{ D(x) }  \left(\partial_x  +\frac{ U'(x)}{2 } \right) \frac{r_n(x)}{K_n e^{ - \frac{ U(x)}{2}}}   
 \nonumber \\
&&  
= \frac{ 1 }{K_n \sqrt { E_n {\hat Z}  }}  \partial_x   \left( e^{U(x)} r_n(x)    \right)
\label{hatrnrn}
\end{eqnarray}
is reminiscent of the mapping of Eq. \eqref{upsilonfp} 
that suggests the choice
\begin{eqnarray}
{\hat r}_n(x)  = - \partial_x   \left( e^{U(x)} r_n(x)    \right)
  \ \ \ \ \ \ \ {\rm if } \ \ K_n  = - \frac{ 1 }{ \sqrt { E_n {\hat Z}  }}
\label{rnKnchoice}
\end{eqnarray}
The inversion using the absorbing boundary conditions for $r_n(0)=0$ reads
\begin{eqnarray}
 r_n(x)    =  -   e^{- U(x)}   \int_0^x dy {\hat r}_n(y)  
\label{rnKnchoiceinversion}
\end{eqnarray}

(ii) the left eigenvector ${\hat l}_n(x)  $ written in terms of the left eigenvector $l_n(x)$
\begin{eqnarray}
 {\hat l}_n ( x )  && 
 = \sqrt {\hat Z} e^{  \frac{ {\hat U} (x) }{2} } { \hat \phi}_n (x )
 =  \sqrt { \frac{{\hat Z} D(x)}{E_n}  } e^{  \frac{ U (x) }{2} } \sqrt{ D(x) }  \left(\partial_x  +\frac{ U'(x)}{2 } \right) K_n e^{ - \frac{ U(x)}{2}} l_n(x)
 \nonumber \\
 && = K_n \sqrt { \frac{{\hat Z} }{E_n}  } e^{ -  U (x) }  D(x)  \partial_x  l_n(x)
\label{hatlnln}
\end{eqnarray}
becomes for the choice of Eq. \eqref{rnKnchoice} for $K_n$
\begin{eqnarray}
{\hat l}_n ( x ) = - \frac{ 1 }{ E_n  } e^{ -  U (x) }  D(x)  \partial_x  l_n(x)
  \ \ \ \ \ \ \ {\rm if } \ \ K_n  = - \frac{ 1 }{ \sqrt { E_n {\hat Z}  }}
\label{lnKnchoice}
\end{eqnarray}
The inversion using the absorbing boundary conditions for $l_n(0)=0$ reads
\begin{eqnarray}
 l_n(x) = - E_n \int_0^x  {\hat l}_n ( y ) \frac{  e^{   U (y) } }{ D(y) }    
 =  - E_n \int_0^x  {\hat l}_n ( y )   e^{  -{\hat U} (y) } 
\label{lnKnchoiceinversion}
\end{eqnarray}


\section{ Kemeny convergence times towards the Fokker-Planck steady states }

\label{sec_kemenyFP}

In this section, we analyze the Kemeny convergence times towards the steady states
of the Fokker-Planck dynamics.

\subsection{ Reminder on the Kemeny convergence time ${\hat \tau}_*$ towards the equilibrium steady state ${\hat \rho}_*^{eq}(x)$}

The Kemeny time ${\hat \tau}_*$ 
needed to converge towards the steady state ${\hat \rho}_*^{eq}(x)$
can be defined in two ways (see \cite{us_kemeny} and references therein) 
  \begin{eqnarray}
{\hat \tau}_* ={\hat \tau}^{Spectral}_*= {\hat \tau}^{Space}_* 
\label{kemenyhat}
\end{eqnarray}
The spectral definition corresponds to the sum of the inverses
of the non-vanishing eigenvalues ${\hat E}_n $
 \begin{eqnarray}
{\hat \tau}^{Spectral}_* =  \sum_{n=1}^{+\infty} \frac{1}{{\hat E}_n} 
\label{tauhatspectral}
\end{eqnarray}
while the spatial definition 
leads to the following explicit expressions in terms of spatial integrals involving the dual potential 
${\hat U}(x)$ and the diffusion coefficient $D(x)$
 \begin{eqnarray}
{\hat \tau}^{Space}_* = \frac{1}{ \int_0^L dx' e^{ -{\hat U}(x')}} 
\int_0^L dx \frac{e^{ {\hat U}(x)}}{D(x)  }
\left[ \int_0^x dy e^{ -{\hat U}(y)} \right]
\left[ \int_x^L dz e^{ -{\hat U}(z)} \right]
\label{tauhatspace}
\end{eqnarray}


\subsection{ Translation for the Kemeny convergence time $ \tau_*$ towards the non-equilibrium steady state $ \rho_*^{noneq}(x)$}

Since the excited energies coincide ${\hat E}_n =E_n$ (Eq. \eqref{phinviaQ}),
 the spectral Kemeny time $ \tau^{Spectral}_* $ 
needed to converge towards the non-equilibrium steady state $ \rho_*^{noneq}(x)$
 coincides with Eq. \eqref{tauhatspectral}
 \begin{eqnarray}
 \tau^{Spectral}_*=  \sum_{n=1}^{+\infty} \frac{1}{ E_n}  ={\hat \tau}^{Spectral}_*  
\label{tauspectral}
\end{eqnarray}
So one can translate Eq. \eqref{tauhatspace} by replacing the dual potential ${\hat U}(x)  =  - U(x) + \ln D(x) $ of Eq. \eqref{Uxhat} in order to write the Kemeny time in terms of spatial integrals
involving the the original potential $U(x)$ and the diffusion coefficient
 \begin{eqnarray}
\tau_*={\hat \tau}^{Space}_* = \frac{1}{ \int_0^L dx' \frac{ e^{ U(x')} }{D(x') }} 
\int_0^L dx e^{-U(x)}
\left[ \int_0^x dy \frac{ e^{ U(y)} }{D(y) } \right]
\left[ \int_x^L dz \frac{ e^{ U(z)} }{D(z) } \right]
\label{tauspace}
\end{eqnarray}


\subsection{ Example : diffusion coefficient $D(x)=1$ and saw-tooth opposite potentials $U(x)=-{\hat U}(x)$}

\label{subsec_sawtooth}

Let us now focus on the simple example
where the diffusion coefficient is constant $D(x)=1$
while the opposite forces $F(x)=-  {\hat F}(x)$ take only two values $(\pm f)$ and change at the position $(\alpha L)$ parametrized by $\alpha \in ]0,1[$
 \begin{eqnarray}
F(x) = - {\hat F}(x) && =  f \ \ \ \ {\rm for } \ \ 0<x<\alpha L
\nonumber \\
F(x) = - {\hat F}(x) && =  - f \ \ \ \ \ \ {\rm for } \ \  \alpha L<x<L
\label{forcesaw}
\end{eqnarray}
so that the corresponding opposite potentials $U(x)=-  {\hat U}(x)$ display the saw-tooth behaviors
\begin{eqnarray}
U(x)=- {\hat U}(x) && =  - f x \ \ \ \ \ \ \ \ \ \ \ \ \ \ {\rm for } \ \ 0 \leq x \leq \alpha L
\nonumber \\
U(x) =- {\hat U}(x) && = - 2 f \alpha L  + f x \ \ \ {\rm for } \ \ \alpha L \leq x \leq L
 \label{URsawtooth}
\end{eqnarray}
with the three important extremal values 
\begin{eqnarray}
U(0) = {\hat U}(x=0) && =0
\nonumber \\
U(x=\alpha L)= - {\hat U}(x=\alpha L) && = - f \alpha L
\nonumber \\
U(x=L) = - {\hat U}(x=L) && =  f (1-2 \alpha  ) L
 \label{URsawtoothextreme}
\end{eqnarray}

The Kemeny time can be explicitly computed via the real-space expressions
of Eqs \eqref{tauhatspace} and \eqref{tauspace}
and reads \cite{us_kemeny}
 \begin{eqnarray}
\tau_*={\hat \tau}_* && = \frac{  2 e^{f (1-  \alpha ) L  }   
   +  e^{ f (1-2 \alpha) L } [ f (1- 2\alpha) L - 4 ]
 +   e^{ -  f \alpha L  } [ 6  + 2   f  L ]
   + f ( 2\alpha-1) L -4 }
   {f^2  \left[ 1+ e^{f (1- 2 \alpha ) L  }-  2 e^{ -  f \alpha L  } \right] } 
\label{kemeny1dsaw}
\end{eqnarray}
Let us now discuss the scaling of the Kemeny time $\tau_*$
for large size $L$ as a function of the two parameters $(\alpha,f)$.


\subsubsection{ Case $f = -  \vert f \vert <0$ : valley for the dual potential ${\hat U}(x) $ 
and mountain for the original potential $U(x)$}

For $f = -  \vert f \vert <0$, corresponding to a linear valley for the dual potential ${\hat U}(x)$,
the convergence towards the equilibrium steady state ${\hat \rho}_*^{eq}$
is governed by the leading linear behavior in $L$ of the Kemeny time
of Eq. \eqref{kemeny1dsaw} 
\begin{eqnarray}
 \tau_* \opsimeq_{L \to + \infty}  \frac{     e^{   \vert f \vert \alpha L  } [   - 2     \vert f \vert  L ] }
   {f^2  \left[ -  2 e^{   \vert f \vert \alpha L  } \right] } 
   = \frac{L }{  \vert f \vert } 
\label{kemeny1dsawvalley}
\end{eqnarray}
For the boundary-driven original problem, this means that the convergence
towards the non-equilibrium steady state $ \rho_*^{noneq} $
will be rapid and given by the linear scaling Eq. \eqref{kemeny1dsawvalley}
when the original potential $U(x)$ corresponds to a linear mountain.


\subsubsection{ Case $f>0$ and $0 <\alpha< 1 $ : mountain for the dual potential ${\hat U}(x) $ and valley for the original potential $U(x)$ }

When the dual potential ${\hat U}(x) $ corresponds to a mountain separating two valleys,
the convergence towards the equilibrium steady state ${\hat \rho}_*^{eq}$
will be very slow and governed by the smallest of the two exponential barriers in $L$
\begin{eqnarray}
{\hat B}_1(L) \equiv {\hat U}(x=\alpha L)  -{\hat U}(x=0) && =  f \alpha L
 \nonumber \\
{\hat B}_2(L) \equiv  {\hat U}(x=\alpha L)  -{\hat U}(x=L) && =  f (1-\alpha)  L
\label{kemeny1dsaw2barr}
\end{eqnarray}
with the following discussion for the leading exponential behavior in $L$ of the Kemeny time
\begin{eqnarray}
\tau_*&&  \opsimeq_{L \to + \infty}   \frac{2}{f^2} e^{f \alpha L} 
   = \frac{2}{f^2} e^{{\hat B}_1(L)} \ \ \ \ {\rm for } \ \ 0 <\alpha< \frac{1}{2}
\nonumber \\
\tau_*&&  \opsimeq_{L \to + \infty}  = \frac{2}{f^2} e^{f (1-\alpha) L} 
   = \frac{2}{f^2} e^{{\hat B}_2(L)}  \ \ \ \ {\rm for } \ \ \frac{1}{2} <\alpha< 1
\nonumber \\
\tau_*  && \opsimeq_{L \to + \infty} \frac{1}{f^2} e^{\frac{f}{2} L} \ \ \ \ {\rm for } \ \alpha=\frac{1}{2}
\label{kemeny1dsawmountain}
\end{eqnarray}
For the boundary-driven original problem, this means that the convergence
towards the non-equilibrium steady state $ \rho_*^{noneq} $
will be very slow and given by the exponential scaling Eq. \eqref{kemeny1dsawmountain}
when the original potential $U(x)$ corresponds to a linear valley.


\section{ Large deviations for the boundary-driven Fokker-Planck dynamics }

\label{sec_largedevFP}

In this section, we describe the large deviations properties
at various levels for empirical time-averages over a large time-window $T$
for the boundary-driven Fokker-Planck dynamics.
We start with the Level 2.5 that is always explicit as recalled in the Introduction.

\subsection{  Explicit large deviations at Level 2.5 for the empirical current $J_e$ and the empirical density $\rho_e(.) $}

\subsubsection{  Joint probability distribution of the empirical current $J_e$ and the empirical density $\rho_e(.) $}

For the boundary-driven Fokker-Planck dynamics,
the joint distribution of the empirical density $\rho_e(.)$ on the whole interval $x \in [0,L]$
and of the empirical current $ J_e$ 
satisfies the large deviation form for large $T$
\begin{eqnarray}
 {\cal P}_T^{[2.5]}[ \rho_e(.),  J_e]   \opsimeq_{T \to +\infty}  
 \delta \left( \rho_e( 0) -c_0  \right)\delta \left( \rho_e( L) -c_L  \right)
e^{- \displaystyle T I_{2.5}[   \rho_e(.),J_e] 
 }
\label{level2.5}
\end{eqnarray}
that involves the constraints on the empirical density at the two boundaries $x=0$ and $x=L$,
while the rate function $I_{2.5}[   \rho_e(.),J_e] $ at level 2.5 that appear in the exponential
reads in terms of the force $F(x)$ or in terms of the potential $U(x)$ of Eq. \eqref{Ux}
\begin{eqnarray}
I_{2.5}[   \rho_e(.),J_e]   &&
=\int_0^L \frac{d  x}{ 4 D ( x) \rho_e( x) } 
\bigg(  J_e  -  \left[ F(x) \rho_e( x) -D(x)   \rho_e'( x) \right] \bigg)^2
\nonumber \\
&&   =
\int_0^L \frac{d  x}{ 4 D ( x) \rho_e( x) } 
\bigg(  J_e  +D ( x) \rho_e( x) \left[  U'( x)+  \frac{  \rho_e'( x) }{\rho_e(x) } \right] \bigg)^2
\label{rate2.5}
\end{eqnarray}

If one expands the square, the linear term in $J_e$ 
can be computed in terms of the boundary conditions 
\begin{eqnarray}
I_{2.5}[   \rho_e(.),J_e]   &&  =
 J_e^2
\int_0^L  \frac{dx}{ 4 D ( x) \rho_e( x) } 
+
\frac{J_e}{2}  \int_0^L d  x  \frac{d}{dx} \left[   U( x)+    \ln (\rho_e( x) ) \right] 
+
\int_0^L dx \frac{ D(x) \rho_e( x)}{ 4   } \left[  U'( x)+ \frac{  \rho_e'( x) }{\rho_e(x) } \right]^2
\nonumber \\
&&    = J_e^2
\int_0^L  \frac{dx}{ 4 D ( x) \rho_e( x) } 
+
\frac{J_e}{2}     \ln \left( \frac{c_L e^{U(L)} }{c_0  } \right) 
+
\int_0^L dx \frac{ D(x) \rho_e( x)}{ 4   } \left[  U'( x)+ \frac{  \rho_e'( x) }{\rho_e(x) } \right]^2
\label{rate2.5expandBC}
\end{eqnarray}

So for any given empirical density $\rho_e(.)$,
the difference between the rate function of Eq. \eqref{rate2.5expandBC}
associated with two opposite values $(\pm J_e)$ of the empirical current
is linear in $J_e$ 
\begin{eqnarray}
I_{2.5}[ \rho_e(.)  ; J_e ] - I_{2.5}[ \rho_e(.)  ; - J_e ]
 = J_e \ln \left( \frac{ c_L e^{U(L)} }{c_0} \right)
\label{rate2.5GC}
\end{eqnarray}
while the coefficient of $J_e$ involves the potential difference $U(L)-U(0)=U(L) $ and 
the boundary conditions $(c_0,c_L)$ imposed by the reservoirs.
This property that characterizes the irreversibility of the dynamics
belongs to the Gallavotti-Cohen fluctuation relations
that have been much studied in the field of non-equilibrium dynamics
(see \cite{galla,kurchan_langevin,Leb_spo,maes1999,jepps,derrida-lecture,harris_Schu,kurchan,searles,zia,maes2009,maes2017,chetrite_thesis,chetrite_HDR} and references therein).


\subsubsection{  Translation of the Level 2.5 for the distribution of the empirical force $F_e(.)$}

It is convenient to introduce the empirical force $F_e(.)$ that would make typical
both the empirical current $J_e$ and the empirical density $\rho_e( .) $
\begin{eqnarray}
F_e(x) \equiv  \frac{J_e +D(x)  \rho_e'(x)    }{\rho_e(x)} 
\label{forceEmpi}
\end{eqnarray}
Using the associated empirical potential 
\begin{eqnarray}
U_e(x) && \equiv - \int_0^x dy \frac{F_e(y)}{D(y)} 
 \label{UxEmpi}
\end{eqnarray}
and the boundary conditions for $\rho_e( x) $, 
one obtains the expressions similar to Eqs \eqref{solusteadycurrent}
and \eqref{steadynoneq} for $J_e$ and $\rho_e(x)$ as a function of $F_e(.)$ using the potential $U_e(.)$
\begin{eqnarray}
   J_e^{[F_e(.)]} && \equiv  \frac{ c_0  - c_Le^{U_e(L) }  }{ \int_0^L dy \frac{ e^{U_e(y) }  } { D(y) } }
\nonumber \\
  \rho_e^{[F_e(.)]}(x) && = e^{-U_e(x)}    \frac{  c_0 \int_x^L dy \frac{ e^{U_e(y) }  } { D(y) } + c_Le^{U_e(L) } \int_0^x dy \frac{ e^{U_e(y) }  } { D(y) } }{ \int_0^L dz \frac{ e^{U_e(z) }  } { D(z) } } 
\label{empiricalrhojUe}
\end{eqnarray}
that coincide with the steady current $J_*$ and the steady state $\rho_*^{noneq}(x)$
when the empirical force $F_e(x)$ coincides with the true force $F(x)$.

The probability distribution of the empirical force $F_e(.)$ is
governed by the rate function 
translated from Eq. \eqref{rate2.5diff}
\begin{eqnarray}
I_{2.5}^{Force}[   F_e(.)]    =\int_0^L dx \frac{\rho_e^{[F_e(.)]}( x) }{ 4 D ( x)  } 
\left[    F_e(x)   - F(x)    \right]^2
\label{rate2.5diff}
\end{eqnarray}


\subsubsection{  Typical Gaussian fluctuations around the steady state $\rho^{noneq}_*(.)$ and the steady current $J_*$ from the Level 2.5 }

If one is interested into small typical fluctuations of order $\frac{1}{\sqrt T}$ around the steady state,
one just needs to plug
\begin{eqnarray}
J_e && = J_* + \frac{ j_e }{\sqrt{T} }
\nonumber \\
 \rho_e(x) && =\rho_*^{noneq}(x)+ \frac{ g_e(x) }{\sqrt{T} }
\label{hatrhoq}
\end{eqnarray}
into Eq. \eqref{rate2.5}
to obtain the Gaussian rate function for the rescaled current $j_e$ and rescaled density $g_e(x)$
\begin{eqnarray}
 I_{2.5}^{Gauss} [ g_e(.),j_e] 
 && \equiv  \lim_{T \to + \infty}
 \left( T   I_{2.5}[   \rho_e(.)=\rho_*^{noneq}(.)+ \frac{ g_e(.) }{\sqrt{T} },J_e=J_* + \frac{ j_e }{\sqrt{T} }]  
 \right) 
\nonumber \\
&& =\int_0^L \frac{d  x}{ 4 D ( x) \rho_*^{noneq}(x) } 
\bigg(  j_e  -  \left[ F(x) g_e( x) -D(x)   g_e'( x) \right] \bigg)^2
\label{rate2.5egauss}
\end{eqnarray}
that will govern the probability of the rescaled fluctuations 
of the empirical local observables obtained from Eq. \eqref{level2.5}
\begin{eqnarray}
  {\hat P}^{[2.5]Gauss}_T [ g_e(.),j_e] 
  \opsimeq_{T \to +\infty}  \delta \left( g_e( 0)   \right)\delta \left( g_e( L)   \right)
e^{ \displaystyle -  I_{2.5}^{Gauss} [ g_e(.),j_e] 
 }
\label{level2.5egauss}
\end{eqnarray}

Similarly, if one is interested into the 
small typical fluctuations of order $\frac{1}{\sqrt T}$ of the empirical force $F_e(x)$ around the true force $F(x)$
\begin{eqnarray}
F_e(x) =F(x)+ \frac{ f_e(x) }{\sqrt{T} }
\label{forceEmpigauss}
\end{eqnarray}
one obtains the Gaussian rate function for the rescaled empirical force $f_e(x)$
from Eq. \eqref{rate2.5diff}
\begin{eqnarray}
 I_{2.5}^{Gauss} [ f_e(.)] 
 && \equiv  \lim_{T \to + \infty}
 \left( T   I_{2.5}[   F_e(.)=F(.)+ \frac{ f_e(.) }{\sqrt{T} }]  
 \right) 
    =\int_0^L dx \frac{ \rho_*^{noneq}( x) }{ 4 D ( x)  }  f_e^2(x)  
\label{rate2.5diffgauss}
\end{eqnarray}


\subsection{ Explicit large deviations at Level 2 for the empirical density $\rho_e(.) $ alone }

For any given empirical density $\rho_e(.)$, the optimization of the rate function 
at Level 2.5 of Eq. \eqref{rate2.5expandBC}
over the empirical current $J_e$ 
\begin{eqnarray}
0=\partial_{J_e} I_{2.5}[   \rho_e(.),J_e]    =  J_e
\int_0^L  \frac{dx}{ 2 D ( x) \rho_e( x) } 
+
\frac{1}{2}     \ln \left( \frac{c_L e^{U(L)} }{c_0  } \right) 
\label{rate2.5diffavecBCderi}
\end{eqnarray}
leads to the following optimal value $J_e^{opt} [ \rho_e(.)] $ as a function of the  optimal value empirical density $\rho_e(.) $
\begin{eqnarray}
  J_e^{opt} [ \rho_e(.)]
=\frac{    \ln \left( \frac{c_0  }{c_L e^{U(L)} } \right) }{\int_0^L  \frac{dx}{  D ( x) \rho_e( x) }  }
\label{Jeopt}
\end{eqnarray}
that can be plugged into Eq. \eqref{rate2.5expandBC}
to obtain the rate function at Level 2
\begin{eqnarray}
I_{2}[   \rho_e(.)]  && = I_{2.5}[   \rho_e(.),J_e^{opt}[ \rho_e(.)]]   
= - \frac{    \ln^2 \left( \frac{c_0  }{c_L e^{U(L)} } \right) }{\int_0^L  \frac{dx}{  D ( x) \rho_e( x) }  }
+
\int_0^L dx \frac{ D(x) \rho_e( x)}{ 4   } \left[  U'( x)+ \frac{  \rho_e'( x) }{\rho_e(x) } \right]^2
\label{rate2}
\end{eqnarray}
or into the form of Eq. \eqref{rate2.5} to obtain the alternative form
\begin{eqnarray}
I_{2}[   \rho_e(.)]  && = I_{2.5}[   \rho_e(.),J_e^{opt}[ \rho_e(.)]]   
=
\int_0^L \frac{d  x}{ 4 D ( x) \rho_e( x) } 
\left(  -\frac{    \ln \left( \frac{c_0  }{c_L e^{U(L)} } \right) }{\int_0^L  \frac{dy}{  D ( y) \rho_e( y) }  }
  +D ( x) \rho_e( x) \left[  U'( x)+  \frac{  \rho_e'( x) }{\rho_e(x) } \right] \right)^2
\label{rate2alter}
\end{eqnarray}


\subsection{  Explicit large deviations for the empirical current $J_e $ alone }

\subsubsection{  Rate function $I(J_e)$ for the empirical current $J_e $
and the corresponding scaled cumulants generating function $\mu(k)$  }

The probability distribution ${\cal P}_T(J_e)$ of the empirical current $J_e $ alone
can be obtained from the integration of the joint probability of Eq. \eqref{level2.5}
over the empirical density $\rho_e(.)$
\begin{eqnarray}
{\cal P}_T(J_e) && =  \int {\cal D} \rho_e(.)  {\cal P}_T^{[2.5]}[ \rho_e(.),  J_e]
 \nonumber \\
 && \opsimeq_{T \to +\infty}  
 \int {\cal D} \rho_e(.) 
 \delta \left( \rho_e( 0) -c_0  \right)\delta \left( \rho_e( L) -c_L  \right)
e^{   - T  I_{2.5}[   \rho_e(.),J_e]  }
 \opsimeq_{ T \to + \infty} e^{- T I(J_e) }
\label{largedevJ}
\end{eqnarray}
So the positive rate function $I(J_e)\geq 0$ 
vanishing only at its minimum corresponding to the steady current $J_*$ 
\begin{eqnarray}
I(J_*) =0 = I'(J_*)
\label{typvanish}
\end{eqnarray}
corresponds to the optimization of 
the rate function $I_{2.5}[   \rho_e(.),J_e] $ over the empirical density $ \rho_e(.) $ satisfying the boundary constraints.

The generating function $Z_T(k) $ of empirical current $J_e$
can be evaluated from Eq. \eqref{largedevJ}
via the Laplace saddle-point method for large $T$
\begin{eqnarray}
Z_T(k) \equiv  \int d J_e \ {\cal P}_T(J_e) \ e^{  T k J_e  }
\opsimeq_{ T\to + \infty} \int d J_2 \ e^{ T \left[ k J_e   - I(J_e) \right] } \opsimeq_{ T \to + \infty} e^{ T \mu(k) }
\label{multifz}
\end{eqnarray}
where the generating function $\mu(k)  $ of the scaled cumulants of the empirical current $J_e$
 corresponds to the Legendre transform of the rate function $I(J_e)$ 
as a consequence of the saddle-point evaluation of the integral in $J_e$ in Eq. \eqref{multifz}
\begin{eqnarray}
\mu(k)  && =  k J_e  - I(J_e) 
\nonumber \\
0 && = k - I'(J_e)
\label{legendre}
\end{eqnarray}
with the reciprocal Legendre transform
\begin{eqnarray}
I(J_e)   && =  k J_e - \mu(k) 
\nonumber \\
0 && = J_e   - \mu'(k) 
\label{legendrereci}
\end{eqnarray}

The generating function $Z_T(k)   $ of Eq. \eqref{multifz}
 can be obtained from the joint probability of Eq. \eqref{level2.5} 
 \begin{eqnarray}
Z_T(k)  
  && = \int dJ_e \int {\cal D} \rho_e(.)  {\cal P}_T^{[2.5]}[ \rho_e(.),  J_e]
   e^{  T k J_e }
\nonumber \\
&&  \opsimeq_{T \to +\infty}  
\int dJ_e \int {\cal D} \rho_e(.) 
 \delta \left( \rho_e( 0) -c_0  \right)\delta \left( \rho_e( L) -c_L  \right)
e^{  \displaystyle T \left( k J_e - I_{2.5}[   \rho_e(.),J_e] \right)
 } \opsimeq_{ T \to + \infty} e^{ T \mu(k) }
\label{gene2.5}
\end{eqnarray}
For large $T$, the evaluation via the saddle-point method means that one needs to optimize the following functional that involves the parameter $k$
\begin{eqnarray}
{\cal F}_k[   \rho_e(.),\rho_e'(.),J_e]&& \equiv I_{2.5}[   \rho_e(.),\rho_e'(.),J_e]  - k J_e   =
\int_0^L \frac{d  x}{ 4 D ( x) \rho_e( x) } 
\left[  J_e  +   D ( x)  \rho_e'( x)  - F(x) \rho_e( x)    \right]^2- k J_e
\nonumber \\
&& =\int_0^L \frac{d  x}{ 4 D ( x)  } 
\left( \frac{ \left[  J_e  +   D ( x)  \rho_e'( x)     \right]^2 }{\rho_e( x)} 
- 2 \left[  J_e  +   D ( x)  \rho_e'( x)  \right] F(x)    
+ F^2(x) \rho_e( x) \right) - k J_e
\label{rate2.5prime}
\end{eqnarray}
over the empirical current $J_e$ and over the empirical density $\rho_e(.)$
to obtain the scaled cumulants generating function $\mu(k)$
from the optimal value ${\cal F}_k^{opt}  $ of this functional
\begin{eqnarray}
\mu(k) = - {\cal F}_k^{opt} 
\label{mukfromoptimal}
\end{eqnarray}


\subsubsection{  Optimization of the functional ${\cal F}_k[   \rho_e(.),\rho_e'(.),J_e] $ 
with respect to the current $J_e$ and to the density $\rho_e(.) $  }

The optimization of Eq. \eqref{rate2.5prime}
with respect to the empirical current $J_e$ yields  
\begin{eqnarray}
0 =  \frac{ \partial  {\cal F}_k[   \rho_e(.),\rho_e'(.),J_e] }{\partial J_e}  &&  =
  \int_0^L \frac{ dx } { 2 D(x)  }
 \left[ \frac{J_e  +   D ( x)  \rho_e'( x)  }{\rho_e(x) } -  F(x)      \right]
  - k  
   \label{lagrangianderij}
\end{eqnarray}
The functional derivatives of Eq. \eqref{rate2.5prime}
 with respect to the empirical density $\rho_e(x)$ and with respect to its derivative $ \rho_e'(x) $
\begin{eqnarray}
  \frac{\partial {\cal F}_k[   \rho_e(.),\rho_e'(.),J_e]}{\partial \rho_e(x)}
  && =   
\frac{ 1 } { 4 D(x) }
 \left[  - \left( \frac{j +D(x)  \rho_e'(x)    }{\rho_e(x)} \right)^2  + F^2(x) \right] 
\nonumber \\
   \frac{\partial {\cal F}_k[   \rho_e(.),\rho_e'(.),J_e]}{\partial \rho_e'(x)} 
&& = \frac{1}{2}   \left[  \frac{J_e +D(x)  \rho_e'(x)    }{\rho_e(x)}  - F(x)    \right]
 \label{derirr}
\end{eqnarray}
lead to the following
Euler-Lagrange equation
for the optimization with respect to the density $\rho_e(.)$ 
\begin{eqnarray}
0 && =   \frac{\partial {\cal F}_k[   \rho_e(.),\rho_e'(.),J_e]}{\partial \rho_e(x)} 
- \frac{d}{d x} \left[   \frac{\partial {\cal F}_k[   \rho_e(.),\rho_e'(.),J_e]}{\partial \rho_e'(x)}  \right]
\nonumber 
\\
&& = 
\frac{ 1 } { 4 D(x) }
 \left[ F^2(x) - \left( \frac{j +D(x)  \rho_e'(x)    }{\rho_e(x)} \right)^2    \right] 
- \frac{1}{2} \frac{d}{d x} \left[   \frac{J_e +D(x)  \rho_e'(x)    }{\rho_e(x)}  - F(x)      \right]
\label{euler}
\end{eqnarray}


\subsubsection{  Rewriting the optimization equations in terms of the empirical force $F_e(.) $  }

In terms of the empirical force $F_e(x)$ of Eq. \eqref{forceEmpi},
the Euler-Lagrange of Eq \ref{euler}
corresponds to the following Riccati differential equation  
\begin{eqnarray}
 \frac{ F_e'(x) }{2} + \frac{  F_e^2(x) }{4D(x)}   =\frac{ F'(x) }{2} + \frac{  F^2(x) }{4D(x)}  
   \label{eulerg}
\end{eqnarray}
i.e. the supersymmetric quantum potential $V(x)$ of Eq. \eqref{vfromu} involving the true force $F(x)$
should remain unchanged when the true force $F(x)$ is replaced by the empirical force $F_e(x)$
\begin{eqnarray}
V_e(x)
  \equiv    \frac{ F_e'(x) }{2} + \frac{  F_e^2(x) }{4D(x)} 
= \frac{ F'(x) }{2} + \frac{  F^2(x) }{4D(x)}   \equiv V(x)
\label{vfromuempi}
\end{eqnarray}

The optimization over $J_e$ of Eq. \eqref{lagrangianderij}
reads in terms of the empirical force $F_e(x)$ of Eq. \eqref{forceEmpi}
and its associated potential $U_e(x)$ of Eq. \eqref{UxEmpi}
\begin{eqnarray}
0 =     \int_0^L \frac{ dx } { 2 D(x)  } \left[ F_e(x) -  F(x)      \right]  - k  
=  \frac{1}{2}  \int_0^L  dx  \left[ - U_e'(x)  +U'(x)  \right]  - k 
=   \frac{U(L)-U_e(L)}{2} -k
   \label{lagrangianderije}
\end{eqnarray}
This equation imposes the 
difference between the empirical potential $U_e(x)$ and the  true potential $U(x)$
at the boundary $x=L$, while $U(0)=0=U_e(0)$.
Note that once the empirical force $F_e(.)$ is known,
the empirical force $J_e$ and the empirical density $\rho_e(x)$ 
are given by Eqs \eqref{empiricalrhojUe} as a function of $F_e(.)$ and the corresponding
potential $U_e(.)$.


\subsubsection{  Scaled cumulants generating function $\mu(k)$
 in terms of the optimal empirical force $F^{opt}_e(.)$ }

The goal is then to compute the optimal value of the functional of Eq. \eqref{rate2.5prime}
as a function of the parameter $k$
\begin{eqnarray}
{\cal F}_k^{opt} &&    =
\int_0^L \frac{d  x}{ 4 D ( x) \rho^{opt}_e( x) } 
\left[  J_e^{opt}  +   D ( x)  \frac{d \rho^{opt}_e( x)}{dx}   - F(x) \rho^{opt}_e( x)    \right]^2- k J_e^{opt}
\nonumber \\
&& =\int_0^L dx \frac{\rho^{opt}_e( x) }{ 4 D ( x)}   
\left[ F_e^{opt}( x)  - F(x)  \right]^2- k J_e^{opt}
\nonumber \\
&& =\int_0^L dx \rho^{opt}_e( x)  
\left[ \frac{[F_e^{opt}( x)]^2}{ 4 D ( x)}   + \frac{F( x)^2}{ 4 D ( x)} -  \frac{F_e^{opt}( x) F(x) }{ 2 D ( x)} \right]- k J_e^{opt}
\label{rate2.5cost}
\end{eqnarray}
One may use Eq. \eqref{eulerg} to replace $\frac{  F^2(x) }{4D(x)} $ to obtain
\begin{eqnarray}
{\cal F}_k^{opt} &&    =
\frac{1}{2} \int_0^L dx \rho^{opt}_e( x)  \frac{d}{dx} 
\left[ F_e^{opt}(x) - F(x)  \right]
+
\int_0^L dx \rho^{opt}_e( x)  
 \frac{F_e^{opt}( x) [F_e^{opt}( x) - F(x) ] }{ 2 D ( x)}    
- k J_e^{opt}
\label{rate2.5costsuite}
\end{eqnarray}
The first integral can be rewritten via an integration by parts using 
the boundary conditions for $\rho^{opt}_e( x) $ and using Eq. \eqref{forceEmpi}
to replace $\frac{d  \rho^{opt}_e( x)}{dx}$
\begin{eqnarray}
&& \frac{1}{2} \int_0^L dx \rho^{opt}_e( x)  \frac{d}{dx} \bigg( F_e^{opt}(x) - F(x)  \bigg)
 = \frac{1}{2}\left[ \rho^{opt}_e( x)  \bigg( F_e^{opt}(x) - F(x)  \bigg)  \right]_{x=0}^{x=L} - \frac{1}{2} \int_0^L dx \left[ F_e^{opt}(x) - F(x)  \right]  \frac{d \rho^{opt}_e( x)}{dx} 
\nonumber \\
&&   =  \frac{ c_L  \bigg( F_e^{opt}(L) - F(L)  \bigg) 
- c_0  \bigg( F_e^{opt}(0) - F(0)  \bigg)}{2}
- \frac{1}{2} \int_0^L dx \left[ F_e^{opt}(x) - F(x)  \right]  \frac{ \rho^{opt}_e( x) F_e^{opt}(x) - J_e^{opt} }{D(x)} 
\label{integrationparts}
\end{eqnarray}
Plugging Eq. \eqref{integparts} into Eq. \eqref{rate2.5costsuite} leads to the cancellation of the second 
integral of Eq. \eqref{rate2.5costsuite}
\begin{eqnarray}
{\cal F}_k^{opt} &&    =
 \frac{ c_L  \bigg( F_e^{opt}(L) - F(L)  \bigg) 
- c_0  \bigg( F_e^{opt}(0) - F(0)  \bigg)}{2}
+J_e^{opt}
\left[   \int_0^L dx  \frac{F_e^{opt}(x) - F(x)  }{ 2 D(x)} 
- k \right]
\label{rate2.5costsuite2}
\end{eqnarray}
The factor of $J_e^{opt} $ vanishes as a consequence Eq. \eqref{lagrangianderije},
so that the optimal value ${\cal F}_k^{opt} $ reduces to the first contribution.

In summary, the scaled cumulants generating function $\mu(k)$ of Eq. \eqref{mukfromoptimal}
\begin{eqnarray}
\mu(k) = - {\cal F}_k^{opt} 
=  \frac{ c_0  \bigg( F_e^{opt}(0) - F(0)  \bigg) -  c_L  \bigg( F_e^{opt}(L) - F(L)  \bigg) }{2}
\label{mukfromoptimalBC}
\end{eqnarray}
involves the boundary values of the optimal empirical force $F_e^{opt}(.) $.
that satisfies the Riccati Eq. \eqref{eulerg}
with the condition of Eq. \eqref{lagrangianderije}.


\subsubsection{ Explicit solution for the optimal empirical force $F_e^{opt}(.) $ }

The change of function
\begin{eqnarray}
  F_e(x) = F(x) + \frac{1}{\omega(x)} 
   \label{omega}
\end{eqnarray}
transforms the first-order non-linear Riccati Eq. \eqref{eulerg} for $F_e(x) $
into the first-order linear equation for $\omega(x)$
\begin{eqnarray}
\frac{1}{2 D(x) } =  \omega'(x) -  \frac{  F(x) }{D(x)}  \omega(x) 
= \omega'(x)   +U'(x) \omega(x) = e^{-U(x)} \frac{d}{dx} \left(  e^{U(x)} \omega(x) \right)
   \label{omegaeq}
\end{eqnarray}
The general solution 
\begin{eqnarray}
 \omega(x)   = e^{-U(x) }  \left[ \omega(0) +\int_0^x dy \frac{e^{U(y)}}{2 D(y) }  \right]
   \label{omegaeqsol}
\end{eqnarray}
involves the integration constant $\omega(0)$
that should be determined by the condition of Eq. \eqref{lagrangianderije} 
\begin{eqnarray}
k && =     \int_0^L \frac{ dx } { 2 D(x)  } \left[ F_e(x) -  F(x)      \right] 
= \int_0^L \frac{ dx } { 2 D(x) \omega(x) } 
   \label{lagrangianderijek}
\end{eqnarray}
One can use Eq. \eqref{omegaeq}
to replace $\frac{1}{2 D(x)\omega(x) } $
\begin{eqnarray}
k && 
= \int_0^L dx \left[ \frac{ \omega'(x)}{ \omega(x)}   +U'(x)  \right]
= \ln \left( \frac{\omega(L)}{\omega(0)}\right) +U(L)
=  \ln \left( \frac{ e^{-U(L) }  \left[ \omega(0) +\int_0^L dx \frac{e^{U(x)}}{2 D(x) }  \right]}{\omega(0)}\right) +U(L)
\nonumber \\
&& =  \ln \left( 1+ \frac{1}{\omega(0) }  \int_0^L dx \frac{e^{U(x)}}{2 D(x) }  \right)
   \label{lagrangianderijekbis}
\end{eqnarray}
that one can invert to obtain $\omega(0) $ as a function of $k$ and the partition function $\hat Z $ of Eq. \eqref{zhat}
\begin{eqnarray}
\omega(0) = \frac{ \int_0^L dx \frac{e^{U(x)}}{2 D(x) }  }{e^k - 1 } 
=  \frac{ \int_0^L dx e^{-{\hat U}(x)} }{ 2 (e^k-1)}
= \frac{ \hat Z }{ 2 (e^k-1)}
   \label{omega0}
\end{eqnarray}
So the solution of Eq. \eqref{omegaeqsol} reads
\begin{eqnarray}
 \omega(x)   = \frac{ e^{-U(x) } }{ 2 (e^k-1) }  \left[ e^k \int_0^x dy \frac{e^{U(y)}}{2 D(y) }  
 + \int_x^L dy \frac{e^{U(y)}}{2 D(y) } \right]
   \label{omegaeqsolexpli}
\end{eqnarray}
with the following value at $x=L$ 
\begin{eqnarray}
 \omega(L)   
 =  e^{-U(L) } \hat Z \frac{  e^k  }{ 2 (e^k-1)}
   \label{omegaL}
\end{eqnarray}


\subsubsection{ 
Explicit results for the scaled cumulants generating function $\mu(k)$ and the rate function $I(J_e)$}

The scaled cumulants generating function $\mu(k)$ of Eq. \eqref{mukfromoptimalBC}
reads used using Eqs \eqref{omega0} and \eqref{omegaL}
\begin{eqnarray}
\mu(k) 
&& =  \frac{ c_0  \bigg( F_e^{opt}(0) - F(0)  \bigg) -  c_L  \bigg( F_e^{opt}(L) - F(L)  \bigg) }{2}
=  - \frac{ c_0  }{2 \omega(0) } -  \frac{ c_L  }{2 \omega(L) }
\nonumber \\
&&  = \frac{ c_0 (e^k-1) + c_L e^{U(L) } (e^{-k} -1) }{\hat Z}   
\label{mukexplicit}
\end{eqnarray}

The series expansion in $k$
involves the steady current $J_*$ at first order in $k$ as it should
\begin{eqnarray}
\mu(k) 
 = k \left( \frac{ c_0 - c_L e^{U(L) }  }{\hat Z} \right)
 +\frac{k^2}{2} \left( \frac{ c_0  + c_L e^{U(L) }  }{\hat Z} \right) +O(k^3)
 \equiv k J_* + \frac{k^2}{2} \sigma^2+O(k^3)
\label{rate2.5costfinexpli}
\end{eqnarray}
while the order $k^2$ leads to the rescaled variance of the empirical current $J_e$
\begin{eqnarray}
\sigma^2 \equiv \frac{ c_0  + c_L e^{U(L) }  }{\hat Z} 
\label{rescaledvariance}
\end{eqnarray}

To compute the rate function $I(J_e)$ from the inverse Legendre transform of Eq. \eqref{legendrereci},
one needs to invert
\begin{eqnarray}
J_e   = \mu'(k) =  \frac{ c_0 e^k - c_L e^{U(L) } e^{-k}  }{\hat Z}   
\label{derimu}
\end{eqnarray}
The rewriting as a second-order equation for $e^k$ leads to the positive solution
\begin{eqnarray}
e^k = \frac{ {\hat Z}  J_e + \sqrt{ {\hat Z}^2  J_e^2 + 4 c_0 c_L e^{U(L)} } }{ 2 c_0}
=  \frac{ 2 c_L e^{U(L)} }{  \sqrt{ {\hat Z}^2  J_e^2 + 4 c_0 c_L e^{U(L)} } - {\hat Z}  J_e }
\label{derimuinvert}
\end{eqnarray}
that can be used to compute the rate function $I(J_e)$ via Eq. \eqref{legendrereci}
\begin{eqnarray}
I(J_e)   && =  k J_e - \mu(k) 
\nonumber \\
&& = J_e \ln \left( \frac{ {\hat Z}  J_e + \sqrt{ {\hat Z}^2  J_e^2 + 4 c_0 c_L e^{U(L)} } }{ 2 c_0}\right) 
- \frac{ c_0 \left[ \frac{ {\hat Z}  J_e + \sqrt{ {\hat Z}^2  J_e^2 + 4 c_0 c_L e^{U(L)} } }{ 2 c_0}-1\right]
 + c_L e^{U(L) } \left[ \frac{  \sqrt{ {\hat Z}^2  J_e^2 + 4 c_0 c_L e^{U(L)} } - {\hat Z}  J_e }{ 2 c_L e^{U(L)}} -1 \right] }{\hat Z}  
 \nonumber \\
&& = J_e \ln \left( \frac{ {\hat Z}  J_e + \sqrt{ {\hat Z}^2  J_e^2 + 4 c_0 c_L e^{U(L)} } }{ 2 c_0}\right) 
+ \frac{ c_0+ c_L e^{U(L) } -  \sqrt{ {\hat Z}^2  J_e^2 + 4 c_0 c_L e^{U(L)} }   }{\hat Z}    
\label{legendrerecifinal}
\end{eqnarray}

The Gallavotti-Cohen symmetry 
for the difference between the rate function of Eq. \eqref{legendrerecifinal}
associated with two opposite values $(\pm J_e)$ of the empirical current
\begin{eqnarray}
I(J_e) - I(-J_e)  
&& 
 = J_e \ln \left( \frac{ {\hat Z}  J_e + \sqrt{ {\hat Z}^2  J_e^2 + 4 c_0 c_L e^{U(L)} } }{ 2 c_0}\right) 
 + J_e \ln \left( \frac{ - {\hat Z}  J_e + \sqrt{ {\hat Z}^2  J_e^2 + 4 c_0 c_L e^{U(L)} } }{ 2 c_0}\right) 
\nonumber \\
&&  = J_e \ln \left( \frac{ c_L e^{U(L)} }{c_0} \right)
\label{legendrerecifinalCG}
\end{eqnarray}
coincides with Eq. \eqref{rate2.5GC} as it should.

The growth of the rate function $I(J_e)$ of Eq. \eqref{legendrerecifinal}
for large currents $J_e \to \pm \infty$
are governed by the leading logarithmic terms
\begin{eqnarray}
I(J_e)   \opsimeq_{ J_e \to \pm \infty}  \vert J_e \vert \ln   \vert J_e \vert
\label{legendrerecifinaltails}
\end{eqnarray}
while the value for zero current $J_e=0$ reduces to
\begin{eqnarray}
I(J_e=0)   = \frac{ c_0+ c_L e^{U(L) } -  \sqrt{  4 c_0 c_L e^{U(L)} }   }{\hat Z}  
=  \frac{ \left[ \sqrt{ c_0} - \sqrt{ c_L e^{U(L) } } \right]^2   }{\hat Z}   
\label{legendrerecifinalzero}
\end{eqnarray}


\section{ Conclusions  }

\label{sec_conclusion}

In this paper, we have revisited the boundary-driven non-equilibrium Markov models of non-interacting particles in one dimension, either in continuous space with the Fokker-Planck dynamics involving an arbitrary force $F(x)$ and an arbitrary diffusion coefficient $D(x)$ in the main text, or in discrete space with the Markov jump dynamics involving arbitrary nearest-neighbor transition rates $w(x \pm 1,x)$
in the Appendices. 

We have first considered the similarity transformations from Markov generators towards quantum supersymmetric Hamiltonians. We have described how the mapping from the boundary-driven non-equilibrium dynamics towards some dual equilibrium dynamics of Ref. \cite{tailleurMapping} can be reinterpreted via the two corresponding quantum Hamiltonians that are supersymmetric partners of each other, with the same energy spectra and with simple relations between their eigenstates. We have analyzed the consequences for the spectral decomposition of the boundary-driven dynamics, and we have given explicit expressions for the Kemeny times needed to converge towards the non-equilibrium steady states. This general framework was illustrated with simple examples.

We have then focused on the large deviations at various levels for empirical time-averaged observables over a large time-window $T$. We have started with the always explicit Level 2.5 concerning the joint distribution of the empirical density and of the empirical flows before considering the contractions towards lower levels. In particular, we have described how the rate function for the empirical current can be explicitly computed via the contraction from the Level 2.5 using the properties of the associated quantum supersymmetric Hamiltonians.

As a final remark, let us mention an interesting issue for the future :
since mappings between boundary-driven non-equilibrium towards equilibrium  
also exist for stochastic models of interacting particles in one dimension, as discussed in detail in \cite{tailleurMapping}, 
it is natural to wonder whether these mappings could be also understood via an underlying supersymmetric structure.


\appendix


\section{ Reminder on boundary-driven non-equilibrium Markov jump dynamics  }

\label{app_boundaryJump}

In this Appendix, we describe the boundary-driven non-equilibrium Markov jump dynamics with arbitrary nearest-neighbor transition rates $w(x \pm 1,x)$
and we recall the mapping towards some dual equilibrium dynamics \cite{tailleurMapping}.

\subsection{ Markov jump non-equilibrium dynamics between two reservoirs fixing the boundary densities  }

\subsubsection{ Master equation involving the transition rates $w(x \pm 1,x)$   }

As in the main text, the reservoirs determine the boundary density at $x=0$ and $x=L$ 
\begin{eqnarray}
  \rho_t( x=0 )   && = c_0
  \nonumber \\
  \rho_t(x=L) && = c_L
\label{reservoirs}
\end{eqnarray}
The dynamics for the $(N-1)$ interior densities $\rho_t(x) $ at the sites $x=1,2,..,L-1$ 
can be written as continuity equations
\begin{eqnarray}
 \partial_t \rho_t(x)    =  J_t\left(x-\frac{1}{2}\right) -  J_t\left(x+\frac{1}{2}\right) \ \ {\rm for} \ \ x=1,..,L-1
\label{master}
\end{eqnarray}
where the $N$ currents $J_t\left(x+\frac{1}{2}\right) $ for $x=0,..,L-1$ associated with the links
\begin{eqnarray}
J_t\left(\frac{1}{2}\right) && = w(1,0) c_0 - w(0,1) \rho_t(1)
\nonumber \\
J_t\left(x+\frac{1}{2}\right) &&= w(x+1,x) \rho_t(x) - w(x,x+1) \rho_t(x+1)
\ \ {\rm for} \ \ x=1,..,L-2
\nonumber \\
J_t\left(L-\frac{1}{2}\right) &&= w(L,L-1) \rho_t(L-1) - w(L-1,L) c_L
\label{jlink}
\end{eqnarray}
involve the transition rates $w(x \pm 1,x) $ from the site $x$ towards its two neighbors $(x \pm 1)$.

In order to make the link with the Fokker-Planck dynamics described in the main text,
it is convenient to introduce the following parametrization of the two transition rates
associated with a given link $\left(x+\frac{1}{2}\right)$
\begin{eqnarray}
 w(x+1,x)  =D\left(x+\frac{1}{2}\right) e^{ \frac{ U(x)-U(x+1) }{2} }
 \nonumber \\
  w(x,x+1)  =D\left(x+\frac{1}{2}\right) e^{  \frac{ U(x+1)-U(x) }{2} }
\label{rates}
\end{eqnarray}
in terms of the diffusion coefficient $D\left(x+\frac{1}{2}\right) $ of the link
and of the potential difference $[U(x+1)-U(x) ]$ 
that can be computed from the product and from the ratio of the two
transition rates
\begin{eqnarray}
D\left(x+\frac{1}{2}\right) && = \sqrt{ w(x+1,x)   w(x,x+1) }
\nonumber \\
U(x+1)-U(x) && = \ln \left( \frac{  w(x,x+1) }{  w(x+1,x) } \right)
\label{ratesInversion}
\end{eqnarray}
Choosing the reference $U(0)=0$ leads to the potential
\begin{eqnarray}
U(x)  = \sum_{y=0}^{x-1}\ln  \left( \frac{  w(y,y+1) }{  w(y+1,y) } \right)
\label{Uxjump}
\end{eqnarray}


\subsubsection{ Reminder on the steady state $\rho_*(x) $ and the steady current $J_*$ as a function of the boundary densities $c_0$ and $c_L$}

The steady State $\rho_*(x)$ of Eq. \eqref{master}
is associated with a steady current $J_*\left(x+\frac{1}{2}\right)=J_*$  
that takes the same value $J_*$ on the $N$ links 
\begin{small}
\begin{eqnarray}
J_* && = w(1,0) c_0 - w(0,1) \rho_*(1) 
= D\left(\frac{1}{2}\right)e^{- \frac{ U(1) }{2} }
 \left[ c_0
-  \frac{ \rho_*(1) }{e^{-U(1)}}  \right]
\label{jlinkJsteady} \\
x=1,..,L-2 :\ J_* &&= w(x+1,x) \rho_*(x) - w(x,x+1) \rho_*(x+1)
=D\left(x+\frac{1}{2}\right)e^{- \frac{ U(x)+U(x+1) }{2} }
 \left[ \frac{ \rho_*(x) }{e^{-U(x)}}
-  \frac{ \rho_*(x+1) }{e^{-U(x+1)}}  \right]
\nonumber \\
J_* &&= w(L,L-1) \rho_*(L-1) - w(L-1,L) c_L
=D\left(L-\frac{1}{2}\right)e^{- \frac{ U(L-1)+U(L) }{2} }
 \left[ \frac{ \rho_*(L-1) }{e^{-U(L-1)}}
-  \frac{c_L }{e^{-U(L)}}  \right]
\nonumber
\end{eqnarray}
\end{small}

So the discussion is similar to Eqs \eqref{steadyeq} and \eqref{steadynoneq} :

(i) if the boundary densities $(c_0,c_L)$ imposed by the reservoirs 
satisfy $c_0  = c_Le^{U(L) }$,
then the steady current vanishes $J_*= 0$,
i.e. detailed-balance is satisfied and the corresponding steady state 
reduces to the Boltzmann equilibrium in the potential $U(x)$
\begin{eqnarray}
  \rho^{eq}_*(x) = c_0 e^{  -U(x) } = c_L  e^{ U(L) -U(x) }\ \ \ {\rm if } \ \  J_* =0
\label{steadyeqjump}
\end{eqnarray}

(ii)  if the boundary densities $(c_0,c_L)$ imposed by the reservoirs satisfy $c_0  \ne c_Le^{U(L) }$,
then the steady current is non-vanishing
\begin{eqnarray}
  J_* =  \frac{ c_0  - c_L e^{U(L) }  }{ \displaystyle \sum_{y=0}^{L-1} \frac{ e^{\frac{ U(y)+U(y+1) }{2} }  } { D\left(y+\frac{1}{2}\right) } }
\label{Jsteadyjump}
\end{eqnarray}
and the corresponding non-equilibrium steady state reads 
\begin{eqnarray}
  \rho^{noneq}_*(x) && = e^{-U(x) } \left[  c_0 - J_* \sum_{y=0}^{x-1} \frac{ e^{\frac{ U(y)+U(y+1) }{2} }  } { D\left(y+\frac{1}{2}\right) } \right]
  \nonumber \\
  && =  \frac{ e^{-U(x)} }{ \displaystyle \sum_{y=0}^{L-1} \frac{ e^{\frac{ U(y)+U(y+1) }{2} }  } { D\left(y+\frac{1}{2}\right) } }
  \left[  c_0 \sum_{y=x}^{L-1} \frac{ e^{\frac{ U(y)+U(y+1) }{2} }  } { D\left(y+\frac{1}{2}\right) }
   +c_L e^{U(L) } \sum_{y=0}^{x-1} \frac{ e^{\frac{ U(y)+U(y+1) }{2} }  } { D\left(y+\frac{1}{2}\right) } \right]
\label{steadynoneqjump}
\end{eqnarray}


\subsection{ Mapping from the non-equilibrium dynamics for $\rho_t(x)$ towards some
dual equilibrium dynamics}

\subsubsection{ Reminder on the mapping of \cite{tailleurMapping} from the original density $\rho_t(x)$ 
towards another density ${\hat \rho}_t\left(x+\frac{1}{2}\right) $ }

The mapping discussed in section 9 of \cite{tailleurMapping}
from the original density $\rho_t( x ) $ towards some new density 
that will be here defined on the links
\begin{eqnarray}
 {\hat \rho}_t \left(x+\frac{1}{2}\right)   = e^{U(x)} \rho_t(x)      - e^{U(x+1)} \rho_t(x+1)  \ \ {\rm for } \ \ x=0,..,L-1
\label{hatrho}
\end{eqnarray}
can be inverted via
\begin{eqnarray}
\sum_{x=0}^{y-1} {\hat \rho}_t \left(x+\frac{1}{2}\right)   =  c_0  - e^{U(y)} \rho_t(y) 
\label{hatinvert}
\end{eqnarray}
so that the corresponding total number ${\hat N}_t $ of particles is 
independent of time
\begin{eqnarray}
{\hat N}_t \equiv \sum_{x=0}^{L-1} {\hat \rho}_t \left(x+\frac{1}{2}\right)   =  c_0  - e^{U(L)} c_L
\label{hatNjump}
\end{eqnarray}

With the parametrization of the rates of Eq. \eqref{rates},
the original current of Eq. \eqref{jlink}
\begin{eqnarray}
J_t\left(x+\frac{1}{2}\right) &&= 
D\left(x+\frac{1}{2}\right) e^{ \frac{ U(x)-U(x+1) }{2} } \rho_t(x) 
- D\left(x+\frac{1}{2}\right) e^{  \frac{ U(x+1)-U(x) }{2} } \rho_t(x+1)
\nonumber \\
&& = D\left(x+\frac{1}{2}\right) e^{ - \frac{ U(x)+U(x+1) }{2} }  {\hat \rho}_t \left(x+\frac{1}{2}\right)   
\label{jlinkvershat}
\end{eqnarray}
is proportional to the new density ${\hat \rho}_t \left(x+\frac{1}{2}\right)  $.
The master equation for the new density of Eq. \eqref{jumphatini}
\begin{eqnarray}
\partial_t {\hat \rho}_t \left(\frac{1}{2}\right)  
 && =  0     - e^{U(1)} \partial_t \rho_t(1)  \equiv  0 -  {\hat J}_t(1)
 \nonumber \\
x=1,..,L-2 : \partial_t {\hat \rho}_t \left(x+\frac{1}{2}\right)  
 && = e^{U(x)} \partial_t \rho_t(x)      - e^{U(x+1)} \partial_t \rho_t(x+1)  \equiv  {\hat J}_t(x) -  {\hat J}_t(x+1)
 \nonumber \\
 \partial_t {\hat \rho}_t \left(L-\frac{1}{2}\right)  
 && = e^{U(L-1)} \partial_t \rho_t(L-1)      - 0  \equiv  {\hat J}_t(L-1) -  0
 \label{jumphatini}
\end{eqnarray}
involves the new current associated with sites
\begin{eqnarray}
 {\hat J}_t(0) && = 0
 \nonumber \\
 {\hat J}_t(x) && =  e^{U(x)} \partial_t \rho_t(x) = e^{U(x)} \left[  J_t\left(x-\frac{1}{2}\right) -  J_t\left(x+\frac{1}{2}\right)\right]
 \ \ \ {\rm for } \ \ \ x=1,..,L-1
 \nonumber \\
{\hat J}_t(L) && = 0 
  \label{jumphatj}
\end{eqnarray}
that vanishes at the two boundaries $x=0$ and $x=L$ in agreement with the conservation of ${\hat N}_t={\hat N}$ in Eq. \eqref{hatNjump}.

Plugging Eq. \eqref{jlinkvershat} into Eq. \eqref{jumphatj}
yields that the new current ${\hat J}_t(x)$ written in terms of the two densities ${\hat \rho}_t \left(x\pm\frac{1}{2}\right) $
\begin{eqnarray}
 {\hat J}_t(x)  && = 
    D\left(x-\frac{1}{2}\right) e^{ \frac{ U(x)-U(x-1) }{2} } {\hat \rho}_t \left(x-\frac{1}{2}\right)
  - D\left(x+\frac{1}{2}\right) e^{ \frac{ U(x)-U(x+1) }{2} }  {\hat \rho}_t \left(x+\frac{1}{2}\right) 
\nonumber \\
&&   \equiv \hat w\left(x+\frac{1}{2},x-\frac{1}{2}\right) {\hat \rho}_t \left(x-\frac{1}{2}\right) 
- \hat w\left(x-\frac{1}{2},x+\frac{1}{2}\right)  {\hat \rho}_t \left(x+\frac{1}{2}\right) 
\label{jumphatjnew}
\end{eqnarray}
corresponds to the new transition rates 
\begin{eqnarray}
\hat w\left(x+\frac{1}{2},x-\frac{1}{2}\right) && = D\left(x-\frac{1}{2}\right) e^{ \frac{ U(x)-U(x-1) }{2} }   
 \equiv {\hat D}(x)  e^{ \frac{ {\hat U}\left(x-\frac{1}{2}\right)- {\hat U}\left(x+\frac{1}{2}\right)}{2} }
  \nonumber \\
 \hat w\left(x-\frac{1}{2},x+\frac{1}{2}\right) && = D\left(x+\frac{1}{2}\right) e^{ \frac{ U(x)-U(x+1) }{2} }
  \equiv {\hat D}(x)  e^{ \frac{ {\hat U}\left(x+\frac{1}{2}\right)- {\hat U}\left(x-\frac{1}{2}\right)}{2} }
\label{rateshat}
\end{eqnarray}
The parametrization analog to Eq. \eqref{rates}
involves the new diffusion coefficient
\begin{eqnarray}
{\hat D}(x) && = \sqrt{ \hat w\left(x+\frac{1}{2},x-\frac{1}{2}\right)
\hat w\left(x-\frac{1}{2},x+\frac{1}{2}\right) } 
=  \sqrt{ D\left(x-\frac{1}{2}\right) e^{ \frac{ U(x)-U(x-1) }{2} } 
 D\left(x+\frac{1}{2}\right) e^{ \frac{ U(x)-U(x+1) }{2} } } 
 \nonumber \\
 && =  \sqrt{ D\left(x-\frac{1}{2}\right) 
 D\left(x+\frac{1}{2}\right) e^{ U(x) - \frac{ U(x-1) + U(x+1) }{2} } }  
\label{hatD}
\end{eqnarray}
while the new potential ${\hat U}(.) $ involves the elementary increment
\begin{eqnarray}
{\hat U}\left(x+\frac{1}{2}\right)- {\hat U}\left(x-\frac{1}{2}\right) 
&& =\ln \left( \frac{\hat w\left(x-\frac{1}{2},x+\frac{1}{2}\right)}{\hat w\left(x+\frac{1}{2},x-\frac{1}{2}\right) } \right)
=\ln \left( \frac{D\left(x+\frac{1}{2}\right) e^{ \frac{ U(x)-U(x+1) }{2} }}{ D\left(x-\frac{1}{2}\right) e^{ \frac{ U(x)-U(x-1) }{2} } } \right)
\nonumber \\
&& = - \frac{ U(x+1)-U(x-1) }{2}
+\ln \left( \frac{D\left(x+\frac{1}{2}\right) }{ D\left(x-\frac{1}{2}\right)  } \right) 
\label{hatUincrement}
\end{eqnarray}
leading to the discrete analog of Eq. \eqref{Uxhat}
\begin{eqnarray}
{\hat U}\left(x+\frac{1}{2}\right) = - \frac{ U(x+1)+U(x) }{2}
+\ln \left( D\left(x+\frac{1}{2}\right)  \right) 
\label{hatU}
\end{eqnarray}


\subsubsection{ Equilibrium steady state ${\hat \rho}_*( x ) $ with vanishing steady current ${\hat J}_* =0$ for the new model }

The new steady current of Eq. \eqref{jumphatj} identically vanishes
\begin{eqnarray}
 {\hat J}_*( x )  =0 
\label{jumphatjsteady}
\end{eqnarray}
Via the change of variables of Eq. \eqref{hatrho}, 
the non-equilibrium steady state $ \rho^{noneq}_*(x) $ of Eq. \eqref{steadynoneqjump}
is mapped onto the equilibrium steady state in the potential ${\hat U}\left(x+\frac{1}{2}\right) $ of Eq. \eqref{hatU} using $J_*$ of Eq. \eqref{Jsteadyjump} and $ {\hat N}= c_0 - e^{U(L)} c_L $
of Eq. \eqref{hatNjump}
\begin{eqnarray}
 {\hat \rho}^{eq}_* \left(x+\frac{1}{2}\right)  && = e^{U(x)} \rho_*(x)      - e^{U(x+1)} \rho_*(x+1)  
 ={\hat N}  \frac{ \frac{ e^{\frac{ U(x)+U(x+1) }{2} }  } { D\left(x+\frac{1}{2}\right) }  }{ \displaystyle \sum_{y=0}^{L-1} \frac{ e^{\frac{ U(y)+U(y+1) }{2} }  } { D\left(y+\frac{1}{2}\right) } }  
 = {\hat N} \frac{ e^{- {\hat U}\left(x+\frac{1}{2}\right)} }{\hat Z}
\label{hatrhoeq}
\end{eqnarray}
where the partition function
\begin{eqnarray}
\hat Z  \equiv \sum_{y=0}^{L-1} e^{- {\hat U}\left(x+\frac{1}{2}\right)}
= \sum_{y=0}^{L-1} \frac{ e^{\frac{ U(y)+U(y+1) }{2} }  } { D\left(y+\frac{1}{2}\right) }
 \label{zhatjump}
 \end{eqnarray}
is the discrete analog of Eq. \eqref{zhat}.


\subsubsection{ Iterated mapping from the new density ${\hat \rho}_t( x ) $ towards the another density ${\hat {\hat \rho}}_t( x ) $}

Let us consider the mapping analogous to Eq. \eqref{hatrho} 
involving the potential $ $ of Eq. \eqref{hatU}
to construct the new density $ {\hat {\hat \rho}}_t( x )  $  associated with the sites $x=1,..,L-1$
\begin{eqnarray}
  {\hat {\hat \rho}}_t( x ) && = e^{{\hat U}\left(x-\frac{1}{2}\right)}  {\hat \rho}_t \left(x-\frac{1}{2}\right) 
       - e^{{\hat U}\left(x+\frac{1}{2}\right)}  {\hat \rho}_t \left(x+\frac{1}{2}\right)
       \nonumber \\
       && = e^{- \frac{ U(x)+U(x-1) }{2}
+\ln \left( D\left(x-\frac{1}{2}\right)  \right)} 
       \left[ e^{U(x-1)} \rho_t(x-1)      - e^{U(x)} \rho_t(x) \right]
  \nonumber \\ && 
       - e^{- \frac{ U(x+1)+U(x) }{2}
+\ln \left( D\left(x+\frac{1}{2}\right)  \right)}  
       \left[ e^{U(x)} \rho_t(x)      - e^{U(x+1)} \rho_t(x+1)\right]
        \nonumber \\
       && =D\left(x-\frac{1}{2}\right) 
       \left[  e^{\frac{ U(x-1)-U(x) }{2}} \rho_t(x-1)      - e^{ \frac{ U(x)-U(x-1) }{2}} \rho_t(x) \right]
  \nonumber \\ && 
       - D\left(x+\frac{1}{2}\right)
       \left[ e^{ \frac{ U(x)-U(x+1) }{2}} \rho_t(x)      - e^{ \frac{ U(x+1)-U(x) }{2}} \rho_t(x+1)\right]
       \nonumber \\
       && = w(x,x-1) \rho_t(x-1) + w(x,x+1) \rho_t(x+1)
- \left[ w(x-1,x) + w(x+1,x) \right] \rho_t(x)  =\partial_t  \rho_t(x) 
\label{doubleupsilonjump}
\end{eqnarray}
that coincides with the time derivative of the original density $\rho_t(x) $
as in Eq. \eqref{doubleuhatrho} concerning the Fokker-Planck dynamics discussed in the main text.

Accordingly, the iteration of the rules of Eq. \eqref{hatD}
and \ref{hatU}
leads to the original diffusion coefficient 
\begin{eqnarray}
 {\hat {\hat D}} \left(x+\frac{1}{2}\right) 
&& =  \sqrt{ {\hat D}(x) {\hat D}(x+1) 
e^{ {\hat U}\left(x+\frac{1}{2}\right) - \frac{ {\hat U}\left(x-\frac{1}{2}\right) + {\hat U}\left(x+\frac{3}{2}\right) }{2} } }  
 =  D\left(x+\frac{1}{2}\right) 
 \label{doublehatD}
\end{eqnarray}
and to the original potential
\begin{eqnarray}
 {\hat {\hat U}} (x) && = - \frac{ {\hat U}\left(x+\frac{1}{2}\right)+{\hat U}\left(x-\frac{1}{2}\right) }{2}
+\ln \left( {\hat D}(x)  \right) 
= U(x)
\label{doublehatU}
\end{eqnarray}


\section{ Interpretation via supersymmetric quantum Hamiltonians on the lattice}

\label{app_susyJump}

The mapping from the boundary-driven non-equilibrium dynamics towards
some dual equilibrium dynamics \cite{tailleurMapping} that has been recalled in the previous
Appendix is reinterpreted in the present Appendix
via the corresponding supersymmetric quantum Hamiltonians on the lattice.

\subsection{ Supersymmetric quantum Hamiltonian $H =Q^{\dagger} Q$ associated with the non-equilibrium dynamics for $\rho_t(x) $ }

The change of variables involving the potential $U(x)$ of Eq. \eqref{Uxjump}
\begin{eqnarray}
\rho_t(x)  =  e^{ - \frac{ U(x)}{2}}  \psi_t(x ) \ \ \ {\rm for } \ \ x=0,1,..,L-1,L
\label{ppsijump}
\end{eqnarray}
changes the boundary conditions of Eq. \eqref{reservoirs} into
\begin{eqnarray}
  \psi_t( x=0 )   && = c_0
  \nonumber \\
  \psi_t(x=L) && = c_L e^{  \frac{ U(L)}{2}} 
\label{reservoirspsi}
\end{eqnarray}
while the master equation of Eqs \eqref{master} and \eqref{jlink} for the density $ \rho_t(x) $
\begin{eqnarray}
\partial_t \rho_t(1) && =w(1,0) c_0 + w(1,2) \rho_t(2)
- \left[ w(0,1) + w(2,1) \right] \rho_t(1)  
\nonumber \\
x=2,..,L-2 : \ \ \partial_t \rho_t(x)  &&  =  w(x,x-1) \rho_t(x-1) + w(x,x+1) \rho_t(x+1)
- \left[ w(x-1,x) + w(x+1,x) \right] \rho_t(x)  
\nonumber \\
 \partial_t \rho_t(L-1)  &&  =  w(L-1,L-2) \rho_t(L-2) + w(L-1,L) c_L
- \left[ w(L-2,L-1) + w(L,L-1) \right] \rho_t(L-1)  
\label{masterfull}
\end{eqnarray}
becomes the euclidean Schr\"odinger equation of Eq. \eqref{schropsi}
 for $\psi_t(x )$
 \begin{eqnarray}
  -  \partial_t \psi_t(1)  && =   H(1,0) c_0+ H(1,2) \psi_t(2) + H(1,1)\psi_t(1 ) 
 \nonumber \\
x=2,..,L-2 : \ \  -  \partial_t \psi_t(x)  && =   H(x,x-1) \psi_t(x -1)+ H(x,x+1) \psi_t(x +1) + H(x,x)\psi_t(x ) 
\nonumber \\
 -  \partial_t \psi_t(L-1)  && =   H(L-1,L-2) \psi_t(L-2)+ H(L-1,L) c_L e^{  \frac{ U(L)}{2}} 
   + H(L-1,L-1)\psi_t(L-1 ) 
\label{tridiagH}
\end{eqnarray}
where the quantum Hamiltonian $H=H^{\dagger}$ is the tridiagonal symmetric matrix
whose matrix elements read in terms of the transition rates of Eq. \eqref{rates}  
\begin{eqnarray}
H(x,x)  && = w(x-1,x)+w(x+1,x)
=D\left(x-\frac{1}{2}\right) e^{  \frac{ U(x)-U(x-1) }{2} }
+D\left(x+\frac{1}{2}\right) e^{ \frac{ U(x)-U(x+1) }{2} }
\nonumber \\
H(x,x-1)  && = -w(x,x-1) e^{ \frac{ U(x)-U(x-1) }{2} }
=-D\left(x-\frac{1}{2}\right) 
\nonumber \\
H(x,x+1)  && =-w(x,x+1) e^{ \frac{ U(x)-U(x+1) }{2} }
=- D\left(x+\frac{1}{2}\right) = H(x+1,x)
\label{Hdiagoff}
\end{eqnarray}
i.e. the off-diagonal elements are determined by the diffusion coefficients $D(.)$ of the links,
while the diagonal element $H(x,x)$ corresponds to the total rate $[w(x-1,x)+w(x+1,x) ]$ out of the site $x$.

This tridiagonal matrix $H$ can be factorized
into the supersymmetric form analog to Eq. \eqref{hsusy}
\begin{eqnarray}
H =     Q^{\dagger} Q
\label{hsusymatrix}
\end{eqnarray}
involving the matrix $Q$ and its adjoint $Q^{\dagger}$ with the non-vanishing matrix elements 
\begin{eqnarray}
Q \left( x+\frac{1}{2} ,x \right) =Q^{\dagger} \left( x, x+\frac{1}{2}  \right)
 && = \sqrt{w(x+1,x) } =  \sqrt{ D\left(x+\frac{1}{2}\right) e^{ \frac{ U(x)-U(x+1) }{2} } }
\nonumber \\
Q \left( x+\frac{1}{2} ,x+1 \right)  = Q^{\dagger} \left( x+1, x+\frac{1}{2}  \right)
&& = - \sqrt{w(x,x+1) } = - \sqrt{ D\left(x+\frac{1}{2}\right) e^{  \frac{ U(x+1)-U(x) }{2} } }
\label{qsusytri}
\end{eqnarray}
with the correspondence between Eq. \eqref{Hdiagoff} and Eq. \eqref{qsusytri}
\begin{eqnarray}
H(x,x)  && = \sum_y Q^{\dagger} \left( x, y+\frac{1}{2}  \right)Q\left( y+\frac{1}{2} ,x \right)= Q^2 \left( x-\frac{1}{2} ,x \right) + Q^2 \left( x+\frac{1}{2} ,x \right)
\nonumber \\
H(x,x+1)  && =  \sum_y Q^{\dagger} \left( x, y+\frac{1}{2}  \right)Q\left( y+\frac{1}{2} ,x+1 \right)
= Q \left( x+\frac{1}{2} ,x \right) Q \left( x+\frac{1}{2} ,x+1 \right)
\label{HdiagoffQ}
\end{eqnarray}

The non-equilibrium steady state $\rho^{noneq}_*(x) $ of Eq. \eqref{steadynoneqjump}
translates via Eq. \eqref{ppsijump} into the quantum state
\begin{eqnarray}
\psi_*^{noneq}(x)= e^{ \frac{ U(x)}{2} }  \rho^{noneq}_*(x) 
 =  \frac{ e^{- \frac{ U(x)}{2} } }{ \displaystyle \sum_{y=0}^{L-1} \frac{ e^{\frac{ U(y)+U(y+1) }{2} }  } { D\left(y+\frac{1}{2}\right) } }
  \left[  c_0 \sum_{y=x}^{L-1} \frac{ e^{\frac{ U(y)+U(y+1) }{2} }  } { D\left(y+\frac{1}{2}\right) }
   +c_L e^{U(L) } \sum_{y=0}^{x-1} \frac{ e^{\frac{ U(y)+U(y+1) }{2} }  } { D\left(y+\frac{1}{2}\right) } \right]
\label{psisteadynoneqjump}
\end{eqnarray}


\subsection{ Supersymmetric quantum Hamiltonian ${\hat H} ={\hat Q}^{\dagger} {\hat Q}$ associated with the equilibrium dynamics for ${ \hat \rho}_t(x) $ }

Similarly, the change of variables 
analogous to Eq. \eqref{ppsijump} for the density ${\hat \rho}_t(x) $
that involves the potential ${\hat U}(x)$ of Eq. \eqref{Uxhat}
\begin{eqnarray}
 {\hat \rho}_t \left(x+\frac{1}{2}\right)   
 =e^{ - \frac{ {\hat U}\left(x+\frac{1}{2}\right)}{2}}  {\hat \psi}_t\left(x+\frac{1}{2}\right)   \ \ {\rm for } \ \ x=0,..,L-1
\label{hatrhojump}
\end{eqnarray}
transforms the master equation of Eq. \eqref{jumphatini} \ref{jumphatjnew}
 for ${\hat \rho}_t \left(x+\frac{1}{2}\right)  $
\begin{eqnarray}
\partial_t {\hat \rho}_t \left(\frac{1}{2}\right)  
 && =  \hat w\left(\frac{1}{2},\frac{3}{2}\right)  {\hat \rho}_t \left(\frac{3}{2}\right)
  - \hat w\left(\frac{3}{2},\frac{1}{2}\right) {\hat \rho}_t \left(\frac{1}{2}\right) 
 \nonumber \\
x=1,..,L-2 : \partial_t {\hat \rho}_t \left(x+\frac{1}{2}\right)  
 && = \hat w\left(x+\frac{1}{2},x-\frac{1}{2}\right) {\hat \rho}_t \left(x-\frac{1}{2}\right) 
 + \hat w\left(x+\frac{1}{2},x+\frac{3}{2}\right)  {\hat \rho}_t \left(x+\frac{3}{2}\right)
\nonumber \\
&& - \left[ \hat w\left(x-\frac{1}{2},x+\frac{1}{2}\right) +\hat w\left(x+\frac{3}{2},x+\frac{1}{2}\right)\right]
 {\hat \rho}_t \left(x+\frac{1}{2}\right) 
 \nonumber \\
 \partial_t {\hat \rho}_t \left(L-\frac{1}{2}\right)  
 && =  \hat w\left(L-\frac{1}{2},L-\frac{3}{2}\right) {\hat \rho}_t \left(L-\frac{3}{2}\right) 
- \hat w\left(L-\frac{3}{2},L-\frac{1}{2}\right)  {\hat \rho}_t \left(L-\frac{1}{2}\right) 
 \label{jumphatinifull}
\end{eqnarray}
into the euclidean Schr\"odinger equation for ${\hat \psi}_t\left(x+\frac{1}{2}\right)$
\begin{eqnarray}
-  \partial_t {\hat \psi}_t\left(\frac{1}{2}\right) && = 
{\hat H} \left(\frac{1}{2},\frac{3}{2}\right){\hat \psi}_t\left(\frac{3}{2}\right)
+ {\hat H} \left(\frac{1}{2},\frac{1}{2}\right){\hat \psi}_t\left(\frac{1}{2}\right)
 \nonumber \\
x=1,..,L-2 : \ \ 
-  \partial_t {\hat \psi}_t\left(x+\frac{1}{2}\right) && = 
{\hat H} \left(x+\frac{1}{2},x-\frac{1}{2}\right){\hat \psi}_t\left(x-\frac{1}{2}\right)
+{\hat H} \left(x+\frac{1}{2},x+\frac{3}{2}\right){\hat \psi}_t\left(x+\frac{3}{2}\right)
\nonumber \\
&& +{\hat H} \left(x+\frac{1}{2},x+\frac{1}{2}\right){\hat \psi}_t\left(x+\frac{1}{2}\right)
 \nonumber \\
 -  \partial_t {\hat \psi}_t\left(L-\frac{1}{2}\right) && = 
{\hat H} \left(L-\frac{1}{2},L-\frac{3}{2}\right){\hat \psi}_t\left(L-\frac{3}{2}\right)
 +{\hat H} \left(L-\frac{1}{2},L-\frac{1}{2}\right){\hat \psi}_t\left(L-\frac{1}{2}\right)
\label{schropsihatjump}
\end{eqnarray}
where the quantum supersymmetric Hermitian Hamiltonian $  {\hat H} = {\hat H}^{\dagger}$
 is the tridiagonal symmetric matrix
with the diagonal in terms of the transition rates of Eq. \eqref{rateshat}  
\begin{eqnarray}
{\hat H} \left(\frac{1}{2},\frac{1}{2}\right) && 
= \hat w\left(\frac{3}{2},\frac{1}{2}\right)
= {\hat D}(1)  e^{ \frac{ {\hat U}\left(\frac{1}{2}\right)- {\hat U}\left(\frac{3}{2}\right)}{2} }
\nonumber \\
x=1,..,L-2 : {\hat H} \left(x+\frac{1}{2},x+\frac{1}{2}\right) && =
\hat w\left(x-\frac{1}{2},x+\frac{1}{2}\right) +\hat w\left(x+\frac{3}{2},x+\frac{1}{2}\right)
\nonumber \\
&& = {\hat D}(x)  e^{ \frac{ {\hat U}\left(x+\frac{1}{2}\right)- {\hat U}\left(x-\frac{1}{2}\right)}{2} }
+  {\hat D}(x+1)  e^{ \frac{ {\hat U}\left(x+\frac{1}{2}\right)- {\hat U}\left(x+\frac{3}{2}\right)}{2} }
\nonumber \\
{\hat H} \left(L-\frac{1}{2},L-\frac{1}{2}\right)  && =
\hat w\left(L-\frac{3}{2},L-\frac{1}{2}\right)
=  {\hat D}(L-1)  e^{ \frac{ {\hat U}\left(L-\frac{1}{2}\right)- {\hat U}\left(L-\frac{3}{2}\right)}{2} }
\label{Hhatdiag}
\end{eqnarray}
while the off-diagonal elements reduce to
\begin{eqnarray}
{\hat H} \left(x+\frac{1}{2},x-\frac{1}{2}\right) = {\hat H} \left(x+\frac{1}{2},x-\frac{1}{2}\right)  && 
=- {\hat D}(x)  
\label{Hhatoff}
\end{eqnarray}

This tridiagonal matrix can be factorized into
\begin{eqnarray}
{\hat H} = {\hat Q}^{\dagger} {\hat Q}
\label{hsusypartnerhatmatrix}
\end{eqnarray}
in terms of the matrix ${\hat Q} $ with the matrix elements
\begin{eqnarray}
{\hat Q} \left( x+1, x+\frac{1}{2}  \right) 
 && = \sqrt{{\hat D}(x+1) e^{ \frac{ {\hat U}\left(x+\frac{1}{2}\right)-{\hat U}\left(x+\frac{3}{2}\right) }{2} }}
 = \sqrt{ D\left(x+\frac{1}{2}\right) e^{  \frac{ U(x+1)-U(x) }{2} } }
 =  - Q^{\dagger} \left( x+1, x+\frac{1}{2} \right)
\nonumber \\
{\hat Q} \left( x, x+\frac{1}{2}  \right)
&& =- \sqrt{{\hat D}(x) e^{  \frac{ {\hat U}\left(x+\frac{1}{2}\right)-{\hat U}\left(x-\frac{1}{2}\right) }{2} }}
= -  \sqrt{ D\left(x+\frac{1}{2}\right) e^{ \frac{ U(x)-U(x+1) }{2} } }
= - Q^{\dagger} \left( x, x+\frac{1}{2}  \right)
\label{partnerqsusytri}
\end{eqnarray}
So the matrix ${\hat Q} $ coincide with the opposite of the matrix $Q^{\dagger}$
\begin{eqnarray}
{\hat Q} &&  =  - Q^{\dagger}
\nonumber \\
{\hat Q}^{\dagger} &&  =  - Q
\label{Qsym}
\end{eqnarray}
As a consequence, the quantum Hamiltonians $H$ and ${\hat H}$ are supersymmetric partners of each other
\begin{eqnarray}
    Q Q^{\dagger} && = {\hat Q}^{\dagger} {\hat Q} = {\hat H}
    \nonumber \\
    {\hat Q} {\hat Q}^{\dagger} && = Q^{\dagger} Q  =  H
\label{hsusypartnermatrix}
\end{eqnarray}
as in Eqs \eqref{hsusypartnerhatdouble} and \eqref{hsusypartnerhat}
concerning the Fokker-Planck dynamics analyzed in the main text,
where the consequences for the relaxation spectra are discussed in detail in section \ref{sec_spectralFP}.


\section{ Kemeny convergence times towards the steady states of the Markov jump dynamics}

\label{app_kemenyJump}

In this Appendix, we analyze the Kemeny convergence times towards the steady states
of the Markov jump dynamics.

\subsection{ Reminder on the Kemeny convergence time ${\hat \tau}_*$ towards the equilibrium steady state ${\hat \rho}_*^{eq}(x)$}

For the Markov jump dynamics of ${\hat \rho}_*^{eq}(x)$
converging towards the equilibrium steady state ${\hat \rho}_*^{eq}(x)$,
the Kemeny convergence time ${\hat \tau}_* ={\hat \tau}^{Spectral}_*= {\hat \tau}^{Space}_*$ of Eq. \eqref{kemenyhat}
can be computed (see \cite{us_kemeny} and references therein)
either from the spectral definition corresponding to the sum of the inverses
of the $(L-1)$ non-vanishing eigenvalues ${\hat E}_n $
 \begin{eqnarray}
{\hat \tau}^{Spectral}_* =  \sum_{n=1}^{L-1} \frac{1}{{\hat E}_n} 
\label{tauhatspectralJump}
\end{eqnarray}
or via the following spatial explicit expression involving the potential 
${\hat U}(x)$ and the diffusion coefficient ${\hat D}(x) $
 \begin{eqnarray}
{\hat \tau}^{Space}_* 
= \frac{1}{ \displaystyle \sum_{ x'=0 }^{L-1}  e^{- {\hat U}\left( x'+\frac{1}{2}\right) }}
\sum_{x=1 }^{L-1} 
 \frac{ e^{ \frac{{\hat U}\left( x-\frac{1}{2}\right)+{\hat U}\left( x+\frac{1}{2}\right)}{2}} }{  {\hat D}(x)  }
\left[ \sum_{ y=0 }^{x-1}  e^{- {\hat U}\left( y+\frac{1}{2}\right) }\right] 
 \left[  \sum_{z=x}^{L-1} e^{- {\hat U}\left( z+\frac{1}{2}\right) } \right]
\label{tauhatspaceJump}
\end{eqnarray}
that corresponds to the discrete analog of Eq. \eqref{tauhatspace}.


\subsection{ Translation for the Kemeny convergence time $ \tau_*$ towards the non-equilibrium steady state $ \rho_*^{noneq}(x)$}

Since the excited energies coincide ${\hat E}_n =E_n$,
 the spectral Kemeny time $ \tau^{Spectral}_* $ 
 needed to converge towards the non-equilibrium steady state $ \rho_*^{noneq}(x)$
 coincides with Eq. \eqref{tauhatspectralJump}
 \begin{eqnarray}
 \tau^{Spectral}_*=  \sum_{n=1}^{L-1} \frac{1}{ E_n}  ={\hat \tau}^{Spectral}_*  
\label{tauspectralJump}
\end{eqnarray}
So one can translate Eq. \eqref{tauhatspaceJump} using Eqs \eqref{hatD}
and \eqref{hatU}
 to write the Kemeny time in terms of the original potential 
$U(x)$ and the original diffusion coefficient $D(x)$
 \begin{eqnarray}
\tau_*={\hat \tau}^{Space}_* = 
&&  \frac{1}{ \displaystyle \sum_{ x'=0 }^{L-1}
  \frac{e^{ \frac{ U(x'+1)+U(x') }{2} } }{ D\left(x'+\frac{1}{2}\right) }}
\sum_{x=1 }^{L-1} e^{-U(x) }
 \left[ \sum_{ y=0 }^{x-1}  \frac{e^{ \frac{ U(y+1)+U(y) }{2} } }{ D\left(y+\frac{1}{2}\right) }\right] 
 \left[  \sum_{z=x}^{L-1} \frac{e^{ \frac{ U(z+1)+U(z) }{2} } }{ D\left(z+\frac{1}{2}\right) } \right]
\label{tauspaceJump}
\end{eqnarray}
that corresponds to the discrete analog of Eq. \eqref{tauspace}.



\section{ Large deviations at various levels for the boundary-driven Markov jump dynamics }

\label{app_largedevJump}

In this Appendix, we describe the large deviations properties
at various levels for empirical time-averages over a large time-window $T$
for the boundary-driven Markov jump dynamics.
We start with the Level 2.5 that is always explicit as recalled in the Introduction.

\subsection{  Explicit large deviations at Level 2.5 }

\subsubsection{  Level 2.5 for the joint probability distribution of the empirical flows $Q_e(x \pm 1,x)$ and the empirical density $\rho_e(x) $}

For the boundary-driven Markov jump dynamics
the joint distribution of the empirical density $\rho_e(x)$ 
and of the empirical flows $ Q_e(x \pm 1,x)$ from $x$ towards $x \pm 1$
satisfies the large deviation form for large $T$
\begin{eqnarray}
&& {\cal P}_T^{[2.5]}[ \rho_e(.) ; Q_e(.,.) ]  \oppropto_{T \to +\infty} 
e^{- T I_{2.5}[ \rho_e(.) ; Q_e(.,.) ] }
\nonumber \\
&&  
\times \delta \left(  \rho_e(0) - c_0 \right) \delta \left(  \rho_e(L) - c_L \right)
 \prod_{x=1}^{L-1} \left[  \left( Q_e(x,x-1) +Q_e(x,x+1) - Q_e(x-1,x) -Q_e(x+1,x)  \right) \right] 
\label{level2.5j}
\end{eqnarray}
that involves, besides the boundary conditions for the empirical density,
 the stationarity constraints for the flows $ Q_e(.,.)$,
i.e. for any $x=1,..,L-1$, 
the total incoming flow $\left[ Q_e(x,x-1) +Q(x,x+1)\right]  $ into $x$ should be equal 
to the total outgoing flow $\left[ Q_e(x-1,x) -Q_e(x+1,x) \right]  $ out of $x$.
The rate function at level 2.5 reads
\begin{eqnarray}
I_{2.5}[ \rho_e(.) ; Q_e(.,.) ]=  \sum_{x=0 }^{L-1} 
\bigg[ && Q_e(x+1,x)  \ln \left( \frac{ Q_e(x+1,x)  }{  w(x+1,x)  \rho_e(x) }  \right)  - Q_e(x+1,x)  + w(x+1,x)  \rho_e(x)
\nonumber \\
&&
+Q_e(x,x+1)  \ln \left( \frac{ Q_e(x,x+1)  }{  w(x,x+1)  \rho_e(x+1) }  \right)  - Q_e(x,x+1)  + w(x,x+1)  \rho_e(x+1)  \bigg]
\label{rate2.5j}
\end{eqnarray}


\subsubsection{  Level 2.5 in terms of the empirical current $J_e$, of the empirical activities
$A_e\left( x+\frac{1}{2}\right)$ and of the empirical density $\rho_e(.) $}

In order to take into account the stationarity constraints for the flows $ Q_e(.,.)$ in Eq. \eqref{level2.5j},
it is convenient to introduce the empirical 
activities $A_e\left( x+\frac{1}{2}\right)$ of the $L$ links corresponding to the sum of the two flows 
$Q_e(x+1,x)$  and $Q_e(x,x+1)$
\begin{eqnarray}
A_e\left( x+\frac{1}{2}\right) \equiv Q_e(x+1,x)  +Q_e(x,x+1)  \ \ \ {\rm for } \ \ x=0,..,L-1
\label{Activity}
\end{eqnarray}
while the empirical current $J_e\left( x+\frac{1}{2}\right)$
corresponding to the difference of the two flows 
$Q_e(x+1,x)$  and $Q_e(x,x+1)$
 should take the same value $J_e$ on all the links in order to 
 satisfy the stationarity constraints for the flows $ Q_e(.,.)$ in Eq. \eqref{level2.5j}
\begin{eqnarray}
J_e \equiv Q_e(x+1,x)  - Q_e(x,x+1)  \ \ \ \ \ \  \ \ \  {\rm for } \ \ x=0,..,L-1
\label{CurrentEmpi}
\end{eqnarray}

Plugging this parametrization of the flows
\begin{eqnarray}
 Q_e(x+1,x)  && = \frac{A_e\left( x+\frac{1}{2}\right) +J_e}{2}
 \nonumber \\
  Q_e(x,x+1)  && =  \frac{A_e\left( x+\frac{1}{2}\right) - J_e}{2}
\label{Flowspara}
\end{eqnarray}
into Eq. \eqref{level2.5j} leads to
the joint distribution of the empirical current $J_e$, of the empirical activities
$A_e\left( x+\frac{1}{2}\right)$ and of the empirical density $\rho_e(.) $
\begin{eqnarray}
{\cal P}_T^{[2.5]}[ \rho_e(.) ; A_e(.) ;J_e] && \oppropto_{T \to +\infty} 
\delta \left(  \rho_e(0) - c_0 \right) \delta \left(  \rho_e(L) - c_L \right)
  e^{- T I_{2.5}[ \rho_e(.) ; A_e(.) ;J_e] }
\label{level2.5aj}
\end{eqnarray}
where the rate function translated from Eq. \eqref{rate2.5j}
 reads
\begin{eqnarray}
I_{2.5}[ \rho_e(.)  ; A_e(.) ;J_e ]=  \sum_{x=0 }^{L-1} 
\bigg[ && \frac{A_e\left( x+\frac{1}{2}\right) }{2}  
\ln \left( \frac{ A_e^2\left( x+\frac{1}{2}\right) -J_e^2  }{4  w(x+1,x)  \rho_e(x)w(x,x+1)  \rho_e(x+1) }  \right) 
\nonumber \\
&&
+ \frac{J_e}{2}  \ln \left( \frac{ \left[ A_e\left( x+\frac{1}{2}\right) +J_e \right] w(x,x+1)  \rho_e(x+1)}
{ \left[A_e\left( x+\frac{1}{2}\right) - J_e \right]  w(x+1,x)  \rho_e(x) }  \right) 
\nonumber \\
&& - A_e\left( x+\frac{1}{2}\right)  + w(x+1,x)  \rho_e(x) + w(x,x+1)  \rho_e(x+1)  \bigg] \ \ \ 
\label{rate2.5aj}
\end{eqnarray}

For any given empirical density $\rho_e(.)$ and empirical activities $A_e(.)$,
the difference between the rate function of Eq. \eqref{rate2.5aj}
associated with two opposite values $(\pm J_e)$ of the empirical current
is linear in $J_e$, while the coefficient can be rewritten 
in terms of the potential $U(x)$ using Eq. \eqref{rates} and in terms of the boundary conditions
for the empirical density $\rho_e(.)$
\begin{eqnarray}
I_{2.5}[ \rho_e(.)  ; A_e(.) ;J_e ] - I_{2.5}[ \rho_e(.)  ; A_e(.) ;- J_e ]
&& = J_e \sum_{x=0 }^{L-1} 
  \ln \left( \frac{  w(x,x+1)  \rho_e(x+1)}
{  w(x+1,x)  \rho_e(x) }  \right) 
\nonumber \\
&& = J_e \ln \left( \frac{ c_L e^{U(L)} }{c_0} \right)
\label{rate2.5ajdiffGC}
\end{eqnarray}
so one obtains the same Gallavotti-Cohen symmetry as in Eq. \eqref{rate2.5GC}
concerning the Fokker-Planck dynamics.


\subsubsection{ Links with the empirical rates $w_e(x\pm1,x) $, the empirical diffusion coefficient $D_e(.)$ and the empirical potential $U_e(.)$ }

If one replaces the empirical flows $ Q_e(x \pm 1,x) $ of Eq. \eqref{Flowspara}
by the empirical rates  
\begin{eqnarray}
w_e(x+1,x) && \equiv \frac{ Q_e(x+1,x)  }{    \rho_e(x) }  
=  \frac{A_e\left( x+\frac{1}{2}\right) +J_e}{2  \rho_e(x)}
\nonumber \\
w_e(x,x+1) && \equiv\frac{ Q_e(x,x+1)  }{    \rho_e(x+1) }  
=  \frac{A_e\left( x+\frac{1}{2}\right) - J_e}{2 \rho_e(x+1)}
\label{rateEmpi}
\end{eqnarray}
and if one introduces the parametrization analog to Eq. \eqref{rates} for these empirical rates
\begin{eqnarray}
 w_e(x+1,x) && =D_e\left(x+\frac{1}{2}\right) e^{ \frac{ U_e(x)-U_e(x+1) }{2} }
 \nonumber \\
  w_e(x,x+1) && =D_e\left(x+\frac{1}{2}\right) e^{  \frac{ U_e(x+1)-U_e(x) }{2} }
\label{ratesEmpiPara}
\end{eqnarray}
one obtains 
the empirical diffusion coefficient $D_e\left(x+\frac{1}{2}\right) $ 
and the empirical potential difference $[U_e(x+1)-U_e(x) ]$ 
for each link via the analog of Eqs \eqref{ratesInversion}
\begin{eqnarray}
D_e\left(x+\frac{1}{2}\right) && = \sqrt{ w_e(x+1,x)   w_e(x,x+1) }
= \sqrt{\frac{ A_e^2\left( x+\frac{1}{2}\right) -J_e^2  }{4    \rho_e(x)  \rho_e(x+1) } } 
\nonumber \\
U_e(x)-U_e(x+1) && =\ln  \left( \frac{  w_e(x+1,x) }{  w_e(x,x+1) } \right)
= \ln  \left(\frac{ \left[ A_e\left( x+\frac{1}{2}\right) +J_e \right]  \rho_e(x+1)}
{ \left[A_e\left( x+\frac{1}{2}\right) - J_e \right]  \rho_e(x) }  \right)
\label{ratesInversionEmpi}
\end{eqnarray}
These expressions will be useful later to perform some contractions.


\subsection{ Explicit Level 2.25 for the joint distribution of the empirical density $\rho_e(.) $ and the empirical current $J_e$ }

The optimization of the rate function $I_{2.5}[ \rho_e(.) ; A_e(.) ;J_e ] $ of Eq. \eqref{rate2.5aj} 
over the activity $A_e\left( x+\frac{1}{2}\right) $ 
\begin{eqnarray}
0 && = \frac{ \partial I_{2.5}[ \rho_e(.) ; A_e(.) ;J_e ]}{ \partial A_e\left( x+\frac{1}{2}\right)}
=   \frac{1 }{2}  
\ln \left( \frac{ A_e^2\left( x+\frac{1}{2}\right) -J_e^2  }{4  w(x+1,x)  \rho_e(x)w(x,x+1)  \rho_e(x+1) }  \right) 
\label{rate2.5ajderia}
\end{eqnarray}
leads to the optimal value as a function of the empirical current $J_e$ and the empirical density $ \rho_e(x)$
\begin{eqnarray}
 A_e^{opt}\left( x+\frac{1}{2}\right) = \sqrt{ J_e^2  +4  w(x+1,x)  \rho_e(x)w(x,x+1)  \rho_e(x+1) }  
\label{rate2.5ajderiaopt}
\end{eqnarray}
that can be plugged into Eq. \eqref{rate2.5aj} 
to obtain the explicit rate function $ $ at Level 2.25 
for the joint distribution of the empirical density $\rho_e(.) $ and the empirical current $J_e$.
\begin{eqnarray}
I_{2.25}[ \rho_e(.)  ;J_e ] && =
I_{2.5}[ \rho_e(.)  ; A_e^{opt}(.) ;J_e ]
\nonumber \\
&& =  \sum_{x=0 }^{L-1} 
\bigg[  J_e  \ln \left( \frac{ \left[  \sqrt{ J_e^2  +4  w(x+1,x)  \rho_e(x)w(x,x+1)  \rho_e(x+1) }  +J_e \right] }
{ 2 w(x+1,x)  \rho_e(x) }  \right) 
\nonumber \\
&& -  \sqrt{ J_e^2  +4  w(x+1,x)  \rho_e(x)w(x,x+1)  \rho_e(x+1) }   + w(x+1,x)  \rho_e(x) + w(x,x+1)  \rho_e(x+1)  \bigg] \ \ \ 
\label{rate2.25}
\end{eqnarray}

This rate function displays the same Gallavotti-Cohen symmetry as Eq. \eqref{rate2.5ajdiffGC}
\begin{eqnarray}
I_{2.25}[ \rho_e(.)  ; J_e ] - I_{2.25}[ \rho_e(.)  ; - J_e ]
 = J_e \ln \left( \frac{ c_L e^{U(L)} }{c_0} \right)
\label{rate2.25ajdiffGC}
\end{eqnarray}


\subsection{ Level 2 for the distribution of empirical density $\rho_e(.) $ alone }

The optimization of the rate function $I_{2.25}[ \rho_e(.) ;J_e ] $ of Eq. \eqref{rate2.25} 
over the empirical current $J_e$ reads
\begin{eqnarray}
0 && = \frac{ \partial I_{2.25}[ \rho_e(.) ;J_e ]}{ \partial J_e}
=    \sum_{x=0 }^{L-1} 
  \ln \left( \frac{ \left[  \sqrt{ J_e^2  +4  w(x+1,x)  \rho_e(x)w(x,x+1)  \rho_e(x+1) }  +J_e \right] }
{ 2 w(x+1,x)  \rho_e(x) }  \right) 
\label{rate2.25derij}
\end{eqnarray}
so that the optimal solution $J_e^{opt}[ \rho_e(.)]$ as a function of the empirical density is not as explicit as 
Eq. \eqref{Jeopt} concerning the Fokker-Planck dynamics
when on wishes to compute the rate function at Level 2
\begin{eqnarray}
&& I_{2}[ \rho_e(.)  ] =
I_{2.25}[ \rho_e(.)  ;J_e^{opt}[ \rho_e(.)] ]
\nonumber \\
&& =  \sum_{x=0 }^{L-1} 
\bigg[  w(x+1,x)  \rho_e(x) + w(x,x+1)  \rho_e(x+1)
  -  \sqrt{ \left( J_e^{opt}[ \rho_e(.)] \right)^2  +4  w(x+1,x)  \rho_e(x)w(x,x+1)  \rho_e(x+1) }   \bigg] \ \ \ 
\label{rate2jump}
\end{eqnarray}


\subsection{  Large deviations for the empirical current $J_e $ alone }

As described in the main text around Eq \ref{largedevJ},
it is convenient to analyze the large deviations for the empirical current $J_e $ alone
via the generating function of Eq. \eqref{gene2.5} involving the rate function 
$I_{2.5}[ \rho_e(.); A_e(.) ;  J_e] $ of Eq. \eqref{rate2.5aj}
 \begin{eqnarray}
&& Z_T(k)  
  = \int dJ_e \int {\cal D} \rho_e(.) \int {\cal D} A_e(.) {\cal P}_T^{[2.5]}[ \rho_e(.); A_e(.) ;  J_e]
   e^{  T k J_e }
\nonumber \\
&&  \opsimeq_{T \to +\infty}  
\int dJ_e \int {\cal D} \rho_e(.) \int {\cal D} A_e(.)
 \delta \left( \rho_e( 0) -c_0  \right)\delta \left( \rho_e( L) -c_L  \right)
e^{  \displaystyle T \left( k J_e - I_{2.5}[ \rho_e(.); A_e(.) ;  J_e] \right)
 } \opsimeq_{ T \to + \infty} e^{ T \mu(k) }
\label{gene2.5j}
\end{eqnarray}
In order to compute the scaled cumulants generating function $\mu(k)$
that governs the behavior for large $T$,
one needs to optimize the functional involving the parameter $k$
\begin{eqnarray}
{\cal F}_k[ \rho_e(.); A_e(.) ;  J_e]
 \equiv  - k J_e + && I_{2.5}[   \rho_e(.),A_e(.),J_e] 
\nonumber \\
   =  - k J_e  + \sum_{x=0 }^{L-1} 
&& \bigg[  \frac{A_e\left( x+\frac{1}{2}\right) }{2}  
\ln \left( \frac{ A_e^2\left( x+\frac{1}{2}\right) -J_e^2  }{4  w(x+1,x)  \rho_e(x)w(x,x+1)  \rho_e(x+1) }  \right) 
\nonumber \\
&&
+ \frac{J_e}{2}  \ln \left( \frac{ \left[ A_e\left( x+\frac{1}{2}\right) +J_e \right] w(x,x+1)  \rho_e(x+1)}
{ \left[A_e\left( x+\frac{1}{2}\right) - J_e \right]  w(x+1,x)  \rho_e(x) }  \right) 
\nonumber \\
&& - A_e\left( x+\frac{1}{2}\right)  + w(x+1,x)  \rho_e(x) + w(x,x+1)  \rho_e(x+1)  \bigg] \label{rate2.5primej}
\end{eqnarray}
over all the empirical variables.


\subsubsection{  Optimization of the functional ${\cal F}_k[   \rho_e(.); A_e(.) ; J_e] $ 
with respect to the empirical variables $[   \rho_e(.); A_e(.) ; J_e] $}

The optimization of Eq. \eqref{rate2.5primej} with respect to the activity $ A_e\left( x+\frac{1}{2}\right) $
can be rewritten using Eq. \eqref{ratesInversion} and Eq. \eqref{ratesInversionEmpi}
\begin{eqnarray}
0  = \frac{ \partial {\cal F}_k[ \rho_e(.); A_e(.) ;  J_e]}{ \partial A_e\left( x+\frac{1}{2}\right)}
=   \frac{1 }{2}  
\ln \left( \frac{ A_e^2\left( x+\frac{1}{2}\right) -J_e^2  }{4  w(x+1,x)  \rho_e(x)w(x,x+1)  \rho_e(x+1) }  \right) 
 = \ln \left( \frac{ D_e\left( x+\frac{1}{2}\right)  }{ D\left( x+\frac{1}{2}\right)}  \right) 
\label{lagrangederia}
\end{eqnarray}
to obtain that the optimal empirical diffusion coefficient $D^{opt}_e\left( x+\frac{1}{2}\right) $
coincides with the true diffusion coefficient $D\left( x+\frac{1}{2}\right) $ on each link
\begin{eqnarray}
 D_e^{opt}\left( x+\frac{1}{2}\right)  = D\left( x+\frac{1}{2}\right)
\label{lagrangederiaSol}
\end{eqnarray}

The optimization of Eq. \eqref{rate2.5primej} with respect to the density $ \rho_e(x) $
can be rewritten using Eq. \eqref{rateEmpi} 
\begin{eqnarray}
0   = \frac{ \partial {\cal F}_k[ \rho_e(.); A_e(.) ;  J_e]}{ \partial \rho_e(x)}
&& = -  \frac{A_e\left( x+\frac{1}{2}\right) }{2 \rho_e(x)}  
-  \frac{A_e\left( x-\frac{1}{2}\right) }{2 \rho_e(x)}  
  + w(x+1,x)  + w(x-1,x)   
\nonumber \\
&&  =  -   w_e(x+1,x)  - w_e(x-1,x)
  + w(x+1,x)  + w(x-1,x)
\label{lagrangederirho}
\end{eqnarray}
Since the total rate $[ w(x+1,x)  + w(x-1,x)]$ out of $x$ 
corresponds to the diagonal element $H(x,x)$ of the supersymmetric Hamiltonian of Eq. \eqref{Hdiagoff},
while the diffusion coefficients $D\left(x+\frac{1}{2}\right)$ determine all the off-diagonal elements 
of the supersymmetric Hamiltonian of Eq. \eqref{Hdiagoff},
the optimization of Eqs \eqref{lagrangederiaSol} and \eqref{lagrangederirho} 
can be summarized by requiring
that the optimal empirical supersymmetric Hamiltonian $H_e^{opt}$ coincides
with the true supersymmetric Hamiltonian $H$
\begin{eqnarray}
H_e^{opt} = H
\label{HEmpi}
\end{eqnarray}
as found in Eq. \eqref{vfromuempi} of the main text for the Fokker-Planck dynamics.

The optimization of Eq. \eqref{rate2.5primej} with respect to the current $J_e$
can be rewritten using Eq. \eqref{rateEmpi} and \ref{ratesInversionEmpi}
\begin{eqnarray}
0  && = \frac{ \partial {\cal F}_k[ \rho_e(.); A_e(.) ;  J_e]}{ \partial J_e}
=    - k   + \sum_{x=0 }^{L-1} 
 \frac{1}{2}  \ln \left( \frac{ \left[ A_e\left( x+\frac{1}{2}\right) +J_e \right] w(x,x+1)  \rho_e(x+1)}
{ \left[A_e\left( x+\frac{1}{2}\right) - J_e \right]  w(x+1,x)  \rho_e(x) }  \right) 
\nonumber \\
&& =
  - k   + \frac{1}{2} \sum_{x=0 }^{L-1} 
  \ln \left( \frac{ w_e(x+1,x)  w(x,x+1)}
{ w_e(x,x+1) w(x+1,x)  }  \right) 
=  - k   + \frac{1}{2} \sum_{x=0 }^{L-1} \left( U_e(x)-U_e(x+1) - U(x)+U(x+1)\right)
\nonumber \\
&&=  - k   + \frac{U(L)-U_e(L)}{2} 
\label{lagrangederij}
\end{eqnarray}
so one one obtains the same condition for the difference between the empirical potential $U_e(x)$ and the  true potential $U(x)$ at the boundary $x=L$
as found in Eq. \eqref{lagrangianderije} of the main text for the Fokker-Planck dynamics.


\subsubsection{  Scaled cumulants generating function $\mu(k)$
 in terms of the optimal potential $U^{opt}_e(.)$ }

The functional of Eq. \eqref{rate2.5primej} can be rewritten
in terms of the empirical rates $w_e(x \pm 1,x)$, the empirical diffusion coefficient $D_e(.)$and 
the empirical potential $U_e(.)$
using Eqs \eqref{rates}, \eqref{ratesEmpiPara} and \eqref{ratesInversionEmpi}
\begin{eqnarray}
{\cal F}_k[ \rho_e(.); A_e(.) ;  J_e] 
&&= -k J_e 
+ \frac{J_e}{2}  \sum_{x=0 }^{L-1}  \bigg( [U_e(x)-U_e(x+1)] - [U(x)-U(x+1)] \bigg) 
 +  \sum_{x=0 }^{L-1} 
  A_e\left( x+\frac{1}{2}\right)   
\ln \left( \frac{ D_e\left( x+\frac{1}{2}\right)  }{D\left( x+\frac{1}{2}\right) }  \right) 
 \nonumber \\ 
&& +  \sum_{x=0 }^{L-1}  w(x+1,x)  \rho_e(x) + \sum_{x=1 }^{L}  w(x-1,x)  \rho_e(x)
-  \sum_{x=0 }^{L-1}  w_e(x+1,x)  \rho_e(x) - \sum_{x=1 }^{L}  w_e(x-1,x)  \rho_e(x)
\nonumber \\
&& =  J_e \left[ -k + \frac{U(L)-U_e(L)}{2}  \right]
 +  \sum_{x=0 }^{L-1} 
  A_e\left( x+\frac{1}{2}\right)   
\ln \left( \frac{ D_e\left( x+\frac{1}{2}\right)  }{D\left( x+\frac{1}{2}\right) }  \right) 
 \nonumber \\ 
&& 
+  \sum_{x=1 }^{L-1} \left[ w(x+1,x)+ w(x-1,x)- w_e(x+1,x) - w_e(x-1,x) \right] \rho_e(x)
 \nonumber \\ 
&& +  \left[ w(1,0)- w_e(1,0)\right]  c_0
+\left[ w(L-1,L) - w_e(L-1,L) \right] c_L
\label{lagrangerewrite}
\end{eqnarray}
Using the optimization equations of
Eqs \eqref{lagrangederiaSol}, \eqref{lagrangederirho}
and \eqref{lagrangederij},
one obtains that the optimal value of the functional of Eq. \eqref{lagrangerewrite}
reduces to the boundary terms of the last line
that can be rewritten using Eqs \eqref{rates} and \eqref{ratesEmpiPara} with Eq. \eqref{lagrangederiaSol}
\begin{eqnarray}
{\cal F}_k^{opt} && =  c_0 \left[ w(1,0)- w_e^{opt}(1,0)\right]  
+c_L \left[ w(L-1,L) - w_e^{opt}(L-1,L) \right] 
\nonumber \\
&& = c_0 D\left(\frac{1}{2}\right) \left[  e^{- \frac{ U(1) }{2} }-e^{- \frac{ U_e^{opt}(1) }{2} } \right]  
+c_L D\left(L-\frac{1}{2}\right) \left[  e^{  \frac{ U(L)-U(L-1) }{2} } - e^{  \frac{ U_e^{opt}(L)-U_e^{opt}(L-1) }{2} } \right] 
  \label{fkopt}
\end{eqnarray}

In summary, the scaled cumulants generating function $\mu(k)$ of Eq. \eqref{mukfromoptimal}
reads using the boundary values $U_e(0)=0=U(0)$
and $U_e(L) = U(L)  - 2 k   $ of Eq. \eqref{BCUeL}
\begin{eqnarray}
\mu(k) = - {\cal F}_k^{opt} 
=  c_0 D\left(\frac{1}{2}\right) \left[  e^{- \frac{ U_e^{opt}(1) }{2} } - e^{- \frac{ U(1) }{2} }\right]  
+c_L D\left(L-\frac{1}{2}\right) e^{\frac{U(L)}{2} }\left[   e^{ - \frac{ U_e^{opt}(L-1) }{2} -k } - e^{ - \frac{ U(L-1) }{2} } \right] 
\label{mukoptimalBC}
\end{eqnarray}
where the optimal empirical potential $U_e^{opt}(x)  $ satisfies 
Eq. \eqref{lagrangederirho} that reads using Eqs \eqref{rates} and \eqref{ratesEmpiPara}
\begin{eqnarray}
      D\left(x+\frac{1}{2}\right) e^{ \frac{ U_e(x)-U_e(x+1) }{2} }
        + D\left(x-\frac{1}{2}\right) e^{  \frac{ U_e(x)-U_e(x-1) }{2} }
  = D\left(x+\frac{1}{2}\right) e^{ \frac{ U(x)-U(x+1) }{2} }  
  + D\left(x-\frac{1}{2}\right) e^{  \frac{ U(x)-U(x-1) }{2} }
\label{lagrangederirhoexpli}
\end{eqnarray}
and the boundary condition of Eq. \eqref{lagrangianderije} at $x=L$
\begin{eqnarray}
U_e(L) = U(L)  - 2 k   
\label{BCUeL}
\end{eqnarray}
 while $U(0)=0=U_e(0)$.


\subsubsection{ Explicit solution for the scaled cumulants generating function $\mu(k)$ }

Since Eq. \eqref{lagrangederirhoexpli} is non-linear, it is technically simpler 
to use the equality of the two supersymmetric Hamiltonians $H_e^{opt}$ and $H$
of Eq. \eqref{HEmpi} with the equality of the two diffusion coefficients 
$D_e^{opt}(.)$ and $D(.)$ of Eq. \eqref{lagrangederiaSol} 
in order to identify their zero-energy solution $\psi_0(x) $ 
satisfying given boundary conditions $\psi_0(0)=b_0 $ and $\psi_0(L)=b_L $
\begin{eqnarray}
0=H_e^{opt} \psi_0(x)= H \psi_0(x)
\label{HEmpizero}
\end{eqnarray}
when written via the analog of Eq. \eqref{psisteadynoneqjump}
either with the empirical potential $U_e(x)$ or with the true potential $U(x)$ 
\begin{eqnarray}
\psi_0(x)
&&  = e^{- \frac{ U_e(x)}{2} } 
\left[  \frac{ \displaystyle 
 b_0 \sum_{y=x}^{L-1} \frac{ e^{\frac{ U_e(y)+U_e(y+1) }{2} }  } { D\left(y+\frac{1}{2}\right) }
   +b_L e^{\frac{U_e(L)}{2} } \sum_{y=0}^{x-1} \frac{ e^{\frac{ U_e(y)+U_e(y+1) }{2} }  } { D\left(y+\frac{1}{2}\right) } }
   { \displaystyle \sum_{y=0}^{L-1} \frac{ e^{\frac{ U_e(y)+U_e(y+1) }{2} }  } { D\left(y+\frac{1}{2}\right) } } \right]
\nonumber \\
&& =  e^{- \frac{ U(x)}{2} } 
\left[  \frac{ \displaystyle 
 b_0 \sum_{y=x}^{L-1} \frac{ e^{\frac{ U(y)+U(y+1) }{2} }  } { D\left(y+\frac{1}{2}\right) }
   +b_L e^{ \frac{U(L)}{2} } \sum_{y=0}^{x-1} \frac{ e^{\frac{ U(y)+U(y+1) }{2} }  } { D\left(y+\frac{1}{2}\right) } }
   { \displaystyle \sum_{y=0}^{L-1} \frac{ e^{\frac{ U(y)+U(y+1) }{2} }  } { D\left(y+\frac{1}{2}\right) } } \right]  
   \label{psiidentification}
\end{eqnarray}
Since this identification should hold for any boundary values $(b_0,b_L)$,
one can identify separately the coefficients of $b_0$ 
\begin{eqnarray}
 e^{- \frac{ U_e(x)}{2} } 
\left[  \frac{ \displaystyle 
  \sum_{y=x}^{L-1} \frac{ e^{\frac{ U_e(y)+U_e(y+1) }{2} }  } { D\left(y+\frac{1}{2}\right) }
    }
   { \displaystyle \sum_{y=0}^{L-1} \frac{ e^{\frac{ U_e(y)+U_e(y+1) }{2} }  } { D\left(y+\frac{1}{2}\right) } } \right]
 =  e^{- \frac{ U(x)}{2} } 
\left[  \frac{ \displaystyle 
  \sum_{y=x}^{L-1} \frac{ e^{\frac{ U(y)+U(y+1) }{2} }  } { D\left(y+\frac{1}{2}\right) }
    }
   { \displaystyle \sum_{y=0}^{L-1} \frac{ e^{\frac{ U(y)+U(y+1) }{2} }  } { D\left(y+\frac{1}{2}\right) } } \right]  
   \label{psiidentificationb0}
\end{eqnarray}
and the coefficients of $b_L$
\begin{eqnarray}
 e^{ \frac{ U_e(L)- U_e(x)}{2} } 
\left[  \frac{ \displaystyle 
   \sum_{y=0}^{x-1} \frac{ e^{\frac{ U_e(y)+U_e(y+1) }{2} }  } { D\left(y+\frac{1}{2}\right) } }
   { \displaystyle \sum_{y=0}^{L-1} \frac{ e^{\frac{ U_e(y)+U_e(y+1) }{2} }  } { D\left(y+\frac{1}{2}\right) } } \right]
 =  e^{ \frac{U(L) - U(x)}{2} } 
\left[  \frac{ \displaystyle 
   \sum_{y=0}^{x-1} \frac{ e^{\frac{ U(y)+U(y+1) }{2} }  } { D\left(y+\frac{1}{2}\right) } }
   { \displaystyle \sum_{y=0}^{L-1} \frac{ e^{\frac{ U(y)+U(y+1) }{2} }  } { D\left(y+\frac{1}{2}\right) } } \right]  
   \label{psiidentificationbL}
\end{eqnarray}

It is convenient to introduce the empirical partition function
\begin{eqnarray}
{\hat Z}_e && \equiv \sum_{y=0}^{L-1} \frac{ e^{\frac{ U_e(y)+U_e(y+1) }{2} }  } { D\left(y+\frac{1}{2}\right) }   
   \label{notazz}
\end{eqnarray}
that is the analog of the true partition function $\hat Z$ of Eq. \eqref{zhatjump}.

Using the boundary values $U_e(0)=0=U(0)$
and $U_e(L) = U(L)  - 2 k   $ of Eq. \eqref{BCUeL},
let us write Eqs \eqref{psiidentificationb0} and \eqref{psiidentificationbL}
for the special case $x=1$
\begin{eqnarray}
 e^{- \frac{ U_e(1)}{2} } 
 - \frac{ 1    }
   { D\left(\frac{1}{2}\right) {\hat Z}_e }
 && =  e^{- \frac{ U(1)}{2} } 
- \frac{ 1       }
   { D\left(\frac{1}{2}\right) {\hat Z} } 
   \nonumber \\
 \frac{  e^{-k} }   { {\hat Z}_e } 
&& =   \frac{  1 }   { {\hat Z} } 
   \label{psiidentificationb0x1}
\end{eqnarray}
as well as for the special case $x=L-1$
\begin{eqnarray}
  \frac{  e^{-k}    }
   { {\hat Z}_e } 
&& =    \frac{  1  }
   { {\hat Z} }  
\nonumber \\
 e^{ - \frac{  U_e(L-1)}{2} -k} 
- \frac{     e^{ \frac{U(L)}{2}  - 2 k  }   }
   {D\left(L-\frac{1}{2}\right) {\hat Z}_e} 
&& =  e^{ - \frac{ U(L-1)}{2} } 
 -  \frac{      e^{ \frac{U(L)}{2}}  }
   {D\left(L-\frac{1}{2}\right) {\hat Z} }  
   \label{psiidentificationbLxL}
\end{eqnarray}

So the second Eq \ref{psiidentificationb0x1} and the first Eq. \eqref{psiidentificationbLxL}
coincide and determine ${\hat Z}_e$ of Eq. \eqref{notazz}
in terms of $k$ and ${\hat Z} $
\begin{eqnarray}
{\hat Z}_e = {\hat Z} e^{-k}
   \label{SoluHatZe}
\end{eqnarray}
It can be plugged into the first Eq \ref{psiidentificationb0x1}
to compute the value of the empirical potential $U_e(x=1)$
\begin{eqnarray}
 e^{- \frac{ U_e(1)}{2} } 
 && =  e^{- \frac{ U(1)}{2} } + \frac{ e^{k}-1    }
   { D\left(\frac{1}{2}\right) {\hat Z}  }
   \label{psiidentificationb0x1first}
\end{eqnarray}
and into the second Eq \ref{psiidentificationb0x1}
to compute the value of the empirical potential $U_e(x=L-1)$
\begin{eqnarray}
e^{-k} e^{ - \frac{  U_e(L-1)}{2} }  
&& =  e^{ - \frac{ U(L-1)}{2} } 
+ \frac{     e^{ \frac{U(L)}{2}     } (e^{-k} -1)  }
   {D\left(L-\frac{1}{2}\right) {\hat Z} } 
   \label{psiidentificationbLxLsecond}
\end{eqnarray}

Plugging Eqs \eqref{psiidentificationb0x1first} and \eqref{psiidentificationbLxLsecond}
into Eq. \eqref{mukoptimalBC},
one obtains that the scaled cumulants generating function $\mu(k)$ 
\begin{eqnarray}
\mu(k) 
= \frac{ c_0 (e^k-1) + c_L e^{U(L) } (e^{-k} -1) }{\hat Z}    
\label{mukfromoptimalBCfinal}
\end{eqnarray}
coincides with the result found in Eq. \eqref{mukexplicit} of the main text concerning the Fokker-Planck dynamics : the only difference is thus in the continuous definition of Eq. \eqref{zhat}
or in the discrete definition of Eq. \eqref{zhatjump} for the partition function $\hat Z$.
As a consequence, the rate function $I(J_e)$ 
corresponding to the Legendre transform of $\mu(k)$ of Eq. \eqref{mukfromoptimalBCfinal}
is given by Eq. \eqref{legendrerecifinal} of the main text with the properties of Eqs
\ref{legendrerecifinalCG}
\ref{legendrerecifinaltails}
\ref{legendrerecifinalzero}.

\end{document}